\documentclass[conference]{IEEEtran}
\IEEEoverridecommandlockouts
\usepackage[utf8]{inputenc}
\usepackage[T1]{fontenc}

\usepackage[style=ieee,sorting=none,maxnames=6,minnames=1,backend=biber]{biblatex}
\usepackage{amsmath,amssymb,amsthm}
\usepackage{physics}      
\usepackage{braket}       

\usepackage{array,tabularx,booktabs}
\newcolumntype{P}[1]{>{\raggedright\arraybackslash}p{#1}}

\usepackage[inline]{enumitem}

\usepackage{graphicx}
\usepackage{float}
\usepackage{placeins}
\usepackage{xcolor}
\usepackage{tikz}
\usetikzlibrary{positioning,arrows.meta,calc,shapes,fit,backgrounds}
\usepackage{ragged2e}
\usepackage{microtype}

\usepackage{algorithm}
\usepackage{algpseudocode}

\usepackage{siunitx}
\DeclareSIUnit{\mHa}{\milli\hartree}
\DeclareSIUnit{\qubit}{q}

\usepackage{pifont}
\newcommand{\cmark}{\ding{51}}
\newcommand{\xmark}{\ding{55}}
\newcommand{\flag}{\ding{46}}

\DeclareUnicodeCharacter{00A0}{~}            
\DeclareUnicodeCharacter{2009}{\,}           
\DeclareUnicodeCharacter{202F}{\,}           
\DeclareUnicodeCharacter{207B}{\ensuremath{-}}   
\DeclareUnicodeCharacter{2076}{\ensuremath{^6}}  
\DeclareUnicodeCharacter{2208}{\ensuremath{\in}} 
\DeclareUnicodeCharacter{2264}{\ensuremath{\le}} 
\DeclareUnicodeCharacter{2691}{\flag}            

\usepackage[hidelinks]{hyperref}

\title{MAESTROCUT: A Closed‑Loop Framework for Dynamic, Noise‑Adaptive and Cryptographically Secure Quantum Circuit Cutting on Near‑Term Hardware}

\author{
\IEEEauthorblockN{Samuel Punch}
\IEEEauthorblockA{\textit{School of Computer Science} \\
\textit{University College Cork}\\
Cork, Ireland \\
samuel.punch@ucc.ie}
\and
\IEEEauthorblockN{Krishnendu Guha}
\IEEEauthorblockA{\textit{School of Computer Science} \\
\textit{University College Cork}\\
Cork, Ireland \\
k.guha@ucc.ie}
}

\begin{document}
\maketitle
\begin{abstract}
We present \textsc{MaestroCut}, a \emph{closed-loop} framework for quantum circuit cutting that makes large circuits practical on NISQ hardware by operating on adaptive fragments rather than fixed, one-shot cuts. \textsc{MaestroCut} comprises four co-designed modules that update online: an incremental multilevel-FM partitioner, a topology-aware Gaussian-process shot allocator (\emph{Topo-GP}), an entropy-gated estimator cascade that switches between classical-shadows and MLE, and a Pauli-compatible confidentiality layer (\textsc{PhasePad-OTP}) with IND-CFA security at $\approx$1\% runtime overhead. The system targets scalable, drift-resilient, and confidential execution while respecting end-to-end SLOs.
  
Circuit cutting at scale is impeded by four intertwined issues: (i) exponential sampling overhead from entangling cuts, (ii) time-varying device drift invalidating static calibrations, (iii) topology-agnostic allocation that wastes shots on correlated errors, and (iv) plaintext fragment exposure to semi-honest providers. Static plans and offline calibrations are brittle under queue and noise fluctuations; conversely, heavyweight privacy can jeopardize latency/throughput budgets. \textsc{MaestroCut} addresses these tensions by coupling drift tracking with topology-aware reallocation and by enforcing confidentiality that is compatible with Pauli pipelines and estimator choice.

We close the loop with Kalman tracking and CUSUM triggers for repartition, water-filling over a heavy-hex correlation proxy (Topo-GP), and an entropy/shot-aware cascade. In Tier-1 simulation (five workloads, shared seeds), this reduces variance relative to uniform allocation and suppresses heavy shot tails while aligning cascade decisions with the predicted $H$–$s$ boundary. In Tier-2 emulation under calibrated Baseline/Noisy/Bursty/Adversarial scenarios, jitter remains well within target ($\leq 150$\,ms; medians \textbf{29/43/101/45 ms}); time-to-first-result meets the \textbf{220 ms} target except under \emph{Noisy} (+1\,ms) and \emph{Adversarial} (+26\,ms) contention (medians \textbf{169/221/169/246 ms}); reliability stays within caps (\textbf{success} \textbf{100/100/99.75/98\%}, \textbf{timeouts} \textbf{0/0/0.25/0\%}, \textbf{errors} \textbf{0/0/0/2\%}); and throughput degrades smoothly without tail amplification (\textbf{46.4k/42.6k/41.4k/43.4k} QPS). \textsc{PhasePad-OTP} preserves end-to-end performance at the 1\% setting (relative throughput $\approx$\,\textbf{98–99\%}), providing confidentiality with near-free system cost. Together, these results demonstrate adaptive, topology-aware cutting with auditable overheads suitable for NISQ-era deployments.
\end{abstract}

\begin{IEEEkeywords}
quantum circuit cutting, dynamic partitioning, drift-aware shot allocation, estimator cascade
\end{IEEEkeywords}

\section{Introduction}\label{sec:intro}

\textit{Circuit cutting} lets NISQ hardware execute problems that exceed on-chip qubit counts by running fragments and classically reconstructing global observables. However, practical deployment faces coupled systems challenges: sampling blow-up from cuts, time-varying noise and queue drift, topology heterogeneity, and provider visibility into fragment structure and measurement strings. These factors limit scalability, distort error budgets, and expose sensitive algorithm information even when final outputs are correct. \textbf{We ask: how can cutting be made deployable under live drift while preserving fragment confidentiality and end-to-end budgets? Our answer is \textsc{MaestroCut}, a closed-loop co-design that adapts partitioning, allocation, and estimation while enforcing security and audit policies, with evidence from Tier-1 simulation and Tier-2 emulation.}

\paragraph*{Challenges}
Deployable cutting on shared back ends must:
\begin{enumerate}[label=(\roman*), leftmargin=1.6em, itemsep=2pt]
  \item control sampling overhead at run time under $\mu$s-scale control loops and minute-scale drift,
  \item adapt partitioning to heterogeneous heavy-hex topologies and changing error profiles,
  \item allocate shots online with provable variance contraction and budget compliance (see Section~\ref{sec:methods} for the bound),
  \item preserve confidentiality of fragments and operational metadata with auditable abort semantics,
  \item meet latency, memory, and energy constraints across QPU, TN, CPU, and GPU targets.
\end{enumerate}

\paragraph*{Prior work and limitations}
Point solutions typically address a single axis in isolation. \emph{Overhead/topology:} offline or static partitioners reduce e-bits but assume frozen calibrations~\cite{lowe2023fastcut,chen2022approximatequantumcircuitcutting}. \emph{Estimation:} contraction and stabilising estimators improve robustness but ignore shot routing~\cite{ufrecht2024optimal,dai2023ancilla}. \emph{Allocation:} ShotQC variants reduce variance yet are commonly tuned offline~\cite{chen2024shotqc,basu2023fragqc}. \emph{Privacy/metadata:} blind-compute protocols hide algorithm content but are rarely integrated with cutting or estimator choice~\cite{Fitzsimons2017}, while recent work shows link and scheduler metadata can leak work profiles~\cite{lu2024quantumleak}. As a result, partitioning, allocation, estimator selection, and confidentiality are not co-optimised in a closed loop.

\paragraph*{Motivating example}
A tenant must choose between two cut strategies with e-bit counts $e_1<e_2$. Under a transient hot-spot on a heavy-hex pair, the lower-$e$ option yields higher variance due to local readout drift. A static plan overspends shots on the degraded region and leaks work profile through shot vectors. In contrast, our monitor observes features derived from queue and calibration deltas, \textit{re-partitions} to reduce exposure, and re-allocates shots via a Kalman-updated plan; if an online leakage or variance budget is exceeded, execution aborts and the event is immutably logged (see Figure~\ref{fig:system-overview} and Figure~\ref{fig:maestrocut-threat-model}).

\paragraph*{Our work (\textsc{MaestroCut})}
\textsc{MaestroCut} is a closed-loop co-design that enforces end-to-end budgets while adapting to drift and topology in real time:
\begin{itemize}[leftmargin=1.2em, itemsep=1pt]
  \item \textbf{Maestro-Partition.} Incremental multilevel-FM hypergraph partitioning that re-cuts under drift to minimise e-bits while respecting device topology.
  \item \textbf{ShotQC-Kalman with Topo-GP.} Online variance tracking with Kalman updates and Gaussian-process priors over heavy-hex error landscapes for shot allocation.
  \item \textbf{Estimator Cascade.} An entropy-gated switch among de-randomised shadows, twirled maximum-likelihood, and MCMC-based contraction for stable post-processing.
  \item \textbf{PhasePad-OTP.} Pauli-compatible padding with the quantum one-time pad (QOTP) and authenticated encryption with associated data (AEAD), plus decoy tests and an auditable kill-switch; security goal is indistinguishability under chosen-fragment attack (IND-CFA).
  \item \textbf{Hybrid execution.} A backend-agnostic path that routes Clifford-heavy fragments to TN/CPU/GPU while reserving non-Clifford fragments for the QPU.
\end{itemize}

\paragraph*{Contributions}
\begin{itemize}[leftmargin=1.2em, itemsep=1pt]
  \item \textbf{Metric and policy.} Online variance and leakage budgets with abort semantics and immutable audit events that compose across partitioning, allocation, and estimation.
  \item \textbf{Orchestration.} A practical pipeline coupling Maestro-Partition, ShotQC-Kalman with Topo-GP, PhasePad-OTP, and a monitored kill-switch (procedure in Section~\ref{sec:methods}; overview in Figure~\ref{fig:system-overview}).
  \item \textbf{Security layer.} A threat model and Universal-Composability (UC)-style games for confidentiality, verifiability, and availability with Local-Shadow checks and shot-vector privacy (Section~\ref{sec:security}; Figure~\ref{fig:maestrocut-threat-model}). For clarity, \emph{blind compute} denotes interactive protocols (e.g., UBQC), whereas \emph{fragment encryption} denotes non-interactive Pauli padding applied to cut fragments.
  \item \textbf{Evidence—Tier-1.} Simulation across representative workloads shows variance contraction versus uniform allocation, suppression of heavy shot tails, and estimator choices that align with the predicted entropy/shot decision boundary.
  \item \textbf{Evidence—Tier-2.} Emulation under calibrated noise and queue dynamics demonstrates stable short-term latencies (jitter), satisfaction of reliability caps, smooth throughput degradation under stress without tail amplification, and preservation of system SLOs when confidentiality is enabled; sensitivity appears only in first-result latency (\emph{time-to-first-result}, TTFR) under adversarial queueing.
  \item \textbf{Evaluation scope.} Claims are substantiated with Tier-1 and Tier-2; \emph{Tier-3 hardware runs are in progress and are not used to support deployability claims in this version}.
\end{itemize}

\paragraph*{Paper organisation}
Section~\ref{sec:background} reviews circuit cutting, allocation, estimators, and confidentiality. Section~\ref{sec:security} states the system and threat model. Section~\ref{sec:methods} details partitioning, allocation, estimators, and orchestration. Section~\ref{sec:experiments} reports Tier-1 and Tier-2 results, with Section~\ref{sec:conclusion} discussing limitations and outlook.

\section{Background and Motivation}
\label{sec:background}

Circuit cutting decomposes monolithic quantum circuits into sub-circuits executable on noisy-intermediate-scale (NISQ) hardware, trading \emph{quantum depth} for \emph{classical post-processing}. Foundational baselines---\textbf{wire cutting}~\cite{peng2020wirecutting} and \textbf{CutQC}~\cite{tang2021cutqc}---established the core cluster/fragment model and practical stitching under NISQ constraints. While enabling VQE/QAOA instances beyond device limits, four systemic obstacles persist:

\begin{enumerate}[leftmargin=1.2em,itemsep=2pt,topsep=2pt]
  \item \textbf{Sampling overhead}---%
        With $e$ wire-cuts, naïve cost scales as $\mathcal{O}(2^{e})$.
        \textbf{For six cuts, this implies 64 settings and $>\!10^{6}$ shots on superconducting hardware.}
        Partial solutions each target a single factor:
        non-maximally entangled cuts ($\approx 1.6^{e}$)~\cite{bechtold2023nmecs};
        ILP-guided QRCC removes \textbf{29\%} of cuts on BV-102 and $\approx$\,\textbf{43\%} on random-QAOA-48~\cite{pawar2024qrcc};
        ShotQC reduces \emph{stitched-variance / sampling overhead} by up to \textbf{19$\times$} on evaluated benchmarks~\cite{chen2024shotqc};
        basis-element neglect ($(1.5\!-\!1.7)^{e}$ in small demos)~\cite{chen2023efficient};
        QCS-Shadows lowers samples for high-weight observables~\cite{chen2024qcs};
        \emph{randomised-measurement cutting} achieves $\tilde{O}(4^{k}/\varepsilon^{2})$ sample overhead for $k$ parallel wire-cuts and proves an information-theoretic lower bound $\Omega(2^{k}/\varepsilon^{2})$; demonstrated on structured QAOA instances~\cite{lowe2023fastcut};
        pruning yields \textbf{16$\times$} fewer settings~\cite{li2024case};
        Metropolis reconstruction runs linearly for $\le 2$ cuts~\cite{chen2022approximatequantumcircuitcutting}.
        None handle drift or confidentiality.
  \item \textbf{Time-dependent drift}---%
        Hour-scale swings in $T_{1}/T_{2}$, readout error, and crosstalk invalidate daily calibrations.
  \item \textbf{Topology blindness}---%
        Uniform shot budgets ignore heavy-hex error landscapes, wasting effort on correlated pairs.
        \emph{FragQC} boosts fidelity by \textbf{$\approx$14.8\%} on IBM hardware but remains offline and shot-agnostic~\cite{basu2023fragqc}.
  \item \textbf{Provider visibility}---%
        Raw fragments leak intellectual property. We target \emph{confidential execution} (fragment/metadata secrecy) via the Pauli one-time pad (QOTP): encrypting an $n$-qubit fragment \emph{requires and suffices with $2n$ uniformly random key bits} for information-theoretic secrecy; this bound is optimal. A restricted special case---\emph{real-amplitude product states}---can be hidden with $n$ bits; we do not rely on this case~\cite{ambainis2000pqc}. QOTP provides confidentiality only (no authentication/integrity). By contrast, \emph{blind-compute} protocols such as UBQC achieve blindness through \emph{interactive} MBQC with client-prepared random single-qubit states, hiding the computation up to unavoidable size leakage; these are orthogonal to our \emph{non-interactive} encrypted-fragment setting~\cite{broadbent2009ubqc}. Recent cloud timing/metadata side-channels further motivate fragment/metadata hardening~\cite{lu2024quantumleak}; ancilla-driven verification~\cite{dai2023ancilla} does not integrate cutting.
\end{enumerate}

\textbf{Table~\ref{tab:fragmentation-core} summarises core frameworks across Overhead, Drift, Topology, and non-interactive fragment confidentiality.}

\begin{table}[t]
\centering\footnotesize
\setlength{\tabcolsep}{6pt}
\caption{Core frameworks. A tick ($\cmark$) indicates the work primarily addresses that axis. Overhead = reduces sampling variance/shot complexity or reduces cut count/settings. A dash (—) denotes not applicable (non-fragmentation).}
\label{tab:fragmentation-core}
\begin{tabular}{@{}lcccc@{}}
\toprule
\textbf{Framework} & \textbf{Overhead} & \textbf{Drift} & \textbf{Topo.} & \textbf{Priv.} \\
\midrule
WireCut (Peng'20)~\cite{peng2020wirecutting}          & \xmark & \xmark & \xmark & \xmark  \\
CutQC (ASPLOS'21)~\cite{tang2021cutqc}                & \cmark & \xmark &\cmark & \xmark \\
FastCut (Lowe'23)~\cite{lowe2023fastcut}              & \cmark & \xmark & \xmark & \xmark \\
MLEcut / MLFT (Perlin'21)~\cite{perlin2021mlecut}     & \xmark & \xmark & \xmark & \xmark \\
ShotQC (Chen'24)~\cite{chen2024shotqc}                & \cmark & \xmark & \xmark & \xmark \\
QCS-Shadows (Chen'24)~\cite{chen2024qcs}              & \cmark & \xmark & \xmark & \xmark \\
QRCC (Pawar'24)~\cite{pawar2024qrcc}                  & \cmark & \xmark & \cmark & \xmark \\
MLFM (Burt'25)~\cite{burt2025mlfm}                    & \cmark & \xmark & \cmark & \xmark \\
Adaptive-PEC (Dasgupta'23)~\cite{Dasgupta2023AdaptiveDrift} & —       & \cmark & \xmark & \xmark \\
CaliScalpel (Fang'24)~\cite{Fang2024CaliScalpel}      & —       & \cmark & \xmark & \xmark \\
\midrule
\textbf{MaestroCut (this work)}                        & \cmark & \cmark & \cmark & \cmark \\
\bottomrule
\end{tabular}
\end{table}

\paragraph*{Complementary knitting.}
Complementing pure cutting, \emph{circuit knitting with classical communication} can provably reduce sampling overhead—for $n$
nonlocal CNOTs from $O(9^{n})$ to $O(4^{n})$—and gives closed-form $\gamma$-factors for many Clifford and controlled-rotation
gates; in other gate families, classical communication yields no advantage. These CC-optimal schemes remain static and noise-blind,
contrasting with our runtime drift-tracking and confidentiality aims~\cite{piveteau2022circuitknitting}.

\paragraph*{Key prior work.}
\textbf{Multilevel-FM}~\cite{burt2025mlfm} reduces e-bits by 33\% but assumes frozen calibrations and plaintext data; we add drift awareness and PhasePad-OTP.  
\textbf{BasisNeglect}~\cite{chen2023efficient} drops zero-weight Pauli terms ($(1.5\!-\!1.7)^{e}$ in small demos), orthogonal to our drift-aware allocation and privacy.  
\textbf{QCS-Shadows}~\cite{chen2024qcs} follows the HKP classical-shadows framework~\cite{huang2020classicalshadows}, reducing sample complexity but remaining noise-blind and non-private.  
\textbf{ShotQC}~\cite{chen2024shotqc} achieves up to 19$\times$ overhead reduction but lacks continuous drift tracking and privacy.  
\textbf{Perlin (MLFT)}~\cite{perlin2021mlecut} provides a maximum-likelihood stitching baseline that enforces a valid probability distribution and can outperform direct execution under fixed shot budgets.  
\textbf{Lowe et\,al.}~\cite{lowe2023fastcut} use randomised measurements to cut wires with $\tilde{O}(4^{k}/\varepsilon^{2})$ samples and prove an $\Omega(2^{k}/\varepsilon^{2})$ lower bound; effective on QAOA-like structure.  
\textbf{Chen (MCMC)}~\cite{chen2022approximatequantumcircuitcutting} approximates reconstruction tensors linearly for $\le 2$ cuts; complementary but not partition-optimised.  
\textbf{Li et al.}~\cite{li2024case} exploit sparsity but ignore noise drift and confidentiality.  
\textbf{NMECS}~\cite{bechtold2023nmecs} lowers per-cut sampling by 1.6$\times$ without addressing key limitations.  
\textbf{QRCC}~\cite{pawar2024qrcc} removes \textbf{29–43\%} of cuts but lacks dynamic adaptation.  
\textbf{UBQC}~\cite{broadbent2009ubqc} and \textbf{ADBQC}~\cite{dai2023ancilla} enable verification-centric privacy via interactive MBQC with size leakage; MaestroCut targets \emph{operational} fragment/metadata confidentiality in a non-interactive setting, strengthened by QOTP (2$n$ key bits, optimal)~\cite{ambainis2000pqc} and motivated by cloud leaks~\cite{lu2024quantumleak}.

\medskip
\noindent\textbf{No prior framework co-optimises all four axes.} \textsc{MaestroCut} closes this gap with a cloud-ready, closed-loop stack that dynamically repartitions circuits, reallocates shots under drift, switches estimators on entropy cues, and encrypts fragments via PhasePad-OTP—delivering scalable, drift-resilient, topology-aware confidential distributed quantum computation.


\section{Threat Model \& Security Layer}\label{sec:security}
This section first outlines the \textsc{MaestroCut} system architecture and run-time orchestration (Figure~\ref{fig:system-overview}), then states the threat model, adversary classes, and the security layer with mitigations (Figure~\ref{fig:maestrocut-threat-model}).

\subsection{System Overview}\label{subsec:system_overview}
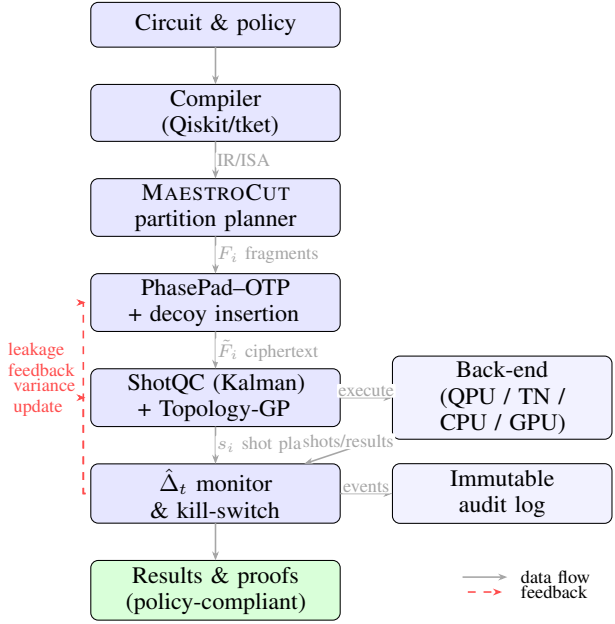
\begin{figure}[H]
  \centering
  \resizebox{0.92\linewidth}{!}{%
  \begin{tikzpicture}[
    font=\small,
    every node/.style={outer sep=0pt},
    box/.style={
      draw, rounded corners=1mm, align=center,
      text width=32mm, minimum height=6mm, inner sep=2pt, fill=blue!10
    },
    side/.style={
      draw, rounded corners=1mm, align=center,
      text width=28mm, minimum height=8mm, inner sep=2pt, fill=blue!5
    },
    outbox/.style={
      draw, rounded corners=1mm, align=center,
      text width=32mm, minimum height=8mm, inner sep=2pt, fill=green!15
    },
    flow/.style={-{Stealth[length=3.5pt]}, semithick, gray!70},
    fb/.style={-{Stealth[length=3.5pt]}, dashed, red!70, semithick,
               shorten >=2pt, shorten <=2pt},
    elab/.style={font=\scriptsize, inner sep=0.6pt, fill=white,
                 rounded corners=0.5pt},
    node distance=5mm
  ]
    \node[box] (client) {Circuit \& policy};
    \node[box, below=of client] (compile) {Compiler \\\ (Qiskit/tket)};
    \node[box, below=of compile] (cut) {\textsc{MaestroCut} \\ partition planner};
    \node[box, below=of cut] (pad) {PhasePad--OTP \\ + decoy insertion};
    \node[box, below=of pad] (alloc) {ShotQC (Kalman) \\ + Topology-GP};
    \node[box, below=of alloc] (mon) {$\hat{\Delta}_t$ monitor \\ \& kill-switch};
    \node[outbox, below=of mon] (out) {Results \& proofs \\ (policy-compliant)};

    \draw[flow] (client) -- (compile);
    \draw[flow] (compile) -- node[elab,right]{IR/ISA} (cut);
    \draw[flow] (cut) -- node[elab,right]{$F_i$ fragments} (pad);
    \draw[flow] (pad) -- node[elab,right]{$\tilde F_i$ ciphertext} (alloc);
    \draw[flow] (alloc) -- node[elab,right]{$s_i$ shot plan} (mon);
    \draw[flow] (mon) -- (out);

    \node[side, right=7mm of alloc] (hw) {Back-end\\(QPU / TN / CPU / GPU)};
    \node[side, right=7mm of mon] (audit) {Immutable\\audit log};

    \draw[flow] (alloc) -- node[elab,above]{execute} (hw);
    \draw[flow] (hw) -- node[elab,above]{shots/results} (mon);
    \draw[flow] (mon) -- node[elab,above]{events} (audit);

    \draw[fb] (mon.west) -- ++(-1mm,0)
      |- node[pos=0.35,left,align=left,font=\scriptsize]{leakage\\feedback}
      (pad.west);

    \draw[fb] (mon.west) -- ++(-1mm,0)
      |- node[pos=0.55,left,align=left,font=\scriptsize]{variance \\ update}
      (alloc.west);

    \node[align=left, anchor=south east] at ($(current bounding box.south east)+(-1mm,1mm)$)
    {\scriptsize \begin{tikzpicture}[baseline={(0,-0.5ex)}]
      \draw[flow] (0,0) -- +(6mm,0); \node[right] at (7mm,0) {data flow};
      \draw[fb]   (0,-2mm) -- +(6mm,0); \node[right] at (7mm,-2mm) {feedback};
    \end{tikzpicture}};

    \node[
      font=\bfseries, anchor=south
    ] at ([yshift=2mm]current bounding box.north)
      {\textsc{MaestroCut} System Overview};
    
  \end{tikzpicture}%
  }
  \caption{System pipeline for \textsc{MaestroCut}: policy-aware compilation, circuit cutting and phase padding, adaptive shot allocation, monitored execution with leakage-bound enforcement, and immutable audit logging.}
  \label{fig:system-overview}
\end{figure}

Figure~\ref{fig:system-overview} shows the end-to-end pipeline.

\paragraph*{Modules and flow.}
\begin{itemize}[leftmargin=1.5em,itemsep=1pt]
  \item \textbf{Circuit \& policy.} The client provides a circuit and policy budgets (variance and leakage).
  \item \textbf{Compiler (Qiskit/tket).} Produces an IR that respects device constraints.
  \item \textbf{Maestro-Partition.} Incremental multilevel-FM hypergraph partitioning; emits fragments $F_i$ while tracking e-bits and cut cost.
  \item \textbf{PhasePad--OTP.} Pauli-compatible padding and encryption of fragments with lightweight decoys for integrity checks.
  \item \textbf{ShotQC (Kalman) + Topology-GP.} Online shot allocation using Kalman-updated variance estimates and Gaussian-process priors over heavy-hex error landscapes.
  \item \textbf{Back-end.} Hybrid execution across QPU / TN / CPU / GPU targets as appropriate.
  \item \textbf{Monitor \& kill-switch.} Observes queue and calibration deltas and enforces policy budgets; aborts on violations and logs the event immutably.
  \item \textbf{Results \& proofs.} Decrypt \& stitch returns observables with audit evidence.
\end{itemize}

\paragraph*{Feedback.}
Two run-time feedback paths (Figure~\ref{fig:system-overview}) ensure adaptivity: (i) leakage feedback into PhasePad to adjust padding and decoys, and (ii) variance updates into ShotQC for budget-respecting reallocation.

\subsection{Threat Model \& Security Layer}\label{subsec:threat_model}
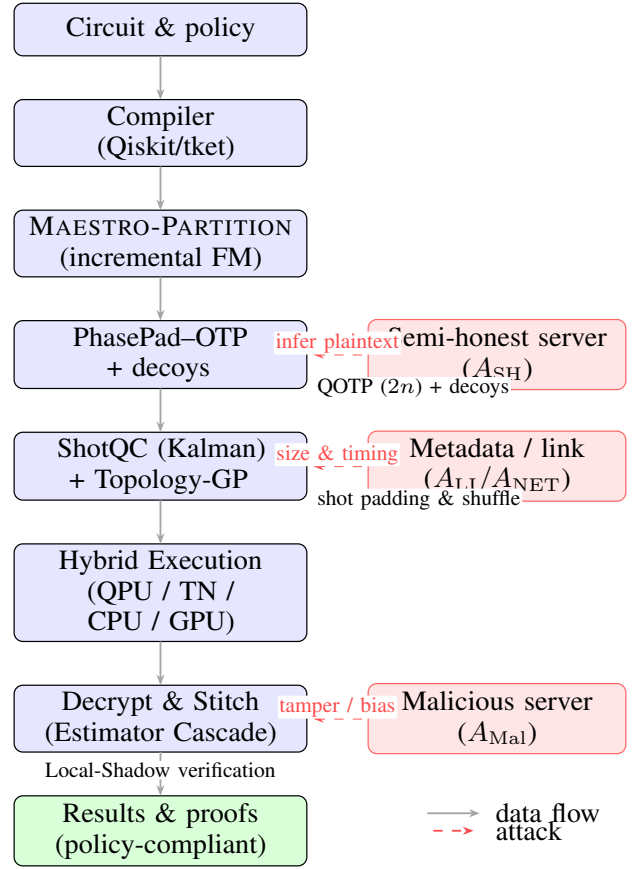
\begin{figure}[H]
  \centering
  \resizebox{0.92\linewidth}{!}{%
\begin{tikzpicture}[
  every node/.style={font=\small, outer sep=0pt},
    box/.style={
      draw, rounded corners=1mm, align=center,
      text width=32mm, minimum height=6mm, inner sep=2pt, fill=blue!10
    },
    outbox/.style={
      draw, rounded corners=1mm, align=center,
      text width=32mm, minimum height=8mm, inner sep=2pt, fill=green!15
    },
    adv/.style={
      draw=red!60, rounded corners=1mm, align=center,
      text width=28mm, minimum height=8mm, inner sep=2pt, fill=red!10
    },
    flow/.style={-{Stealth[length=3.5pt]}, semithick, gray!70},
    atk/.style={-{Stealth[length=3.5pt]}, dashed, red!70, semithick,
                shorten >=2pt, shorten <=2pt},
    elab/.style={font=\scriptsize, inner sep=0.6pt, fill=white,
                 rounded corners=0.5pt},
    node distance=5mm
  ]
    \node[box] (client)  {Circuit \& policy};
    \node[box, below=of client] (compile) {Compiler \\\ (Qiskit/tket)};
    \node[box, below=of compile] (cut)     {\textsc{Maestro-Partition} \\ (incremental FM)};
    \node[box, below=of cut]     (pad)     {PhasePad--OTP \\ + decoys};
    \node[box, below=of pad]     (alloc)   {ShotQC (Kalman) \\ + Topology-GP};
    \node[box, below=of alloc]   (exec)    {Hybrid Execution \\ (QPU / TN / CPU / GPU)};
    \node[box, below=of exec]    (stitch)  {Decrypt \& Stitch \\ (Estimator Cascade)};
    \node[outbox, below=of stitch] (out)   {Results \& proofs \\ (policy-compliant)};

    \draw[flow] (client) -- (compile);
    \draw[flow] (compile) -- (cut);
    \draw[flow] (cut) -- (pad);
    \draw[flow] (pad) -- (alloc);
    \draw[flow] (alloc) -- (exec);
    \draw[flow] (exec) -- (stitch);
    \draw[flow] (stitch) -- (out);

    \node[adv, right=7mm of pad]   (ash)  {Semi-honest server \\ ($A_{\mathrm{SH}}$)};
    \node[adv, right=7mm of alloc] (ali)  {Metadata / link \\ ($A_{\mathrm{LI}}/A_{\mathrm{NET}}$)};
    \node[adv, right=7mm of stitch](amal) {Malicious server \\ ($A_{\mathrm{Mal}}$)};

    \draw[atk] (ash.west)  -- node[elab,above]{infer plaintext} (pad.east);
    \draw[atk] (ali.west)  -- node[elab,above]{size \& timing}  (alloc.east);
    \draw[atk] (amal.west) -- node[elab,above]{tamper / bias}   (stitch.east);

    \node[elab, right=1mm of pad.south east, anchor=west] {QOTP ($2n$) + decoys};
    \node[elab, right=1mm of alloc.south east, anchor=west] {shot padding \& shuffle};
    \node[elab, below=1mm of stitch.south] {Local-Shadow verification};

    \node[align=left, anchor=south east] at ($(current bounding box.south east)+(-1mm,1mm)$)
    {\scriptsize \begin{tikzpicture}[baseline={(0,-0.5ex)}]
    \draw[flow] (0,0) -- +(6mm,0); \node[right] at (7mm,0) {data flow};
    \draw[atk]  (0,-2mm) -- +(6mm,0); \node[right] at (7mm,-2mm) {attack};
    \end{tikzpicture}};

    \node[font=\bfseries, anchor=south]
      at ([yshift=2mm]current bounding box.north)
      {\textsc{MaestroCut} Threat Model};

  \end{tikzpicture}%
  }
  \caption{Threat model over the \textsc{MaestroCut} pipeline. Red boxes denote adversaries; dashed arrows indicate attack surfaces; inline labels show primary mitigations.}
  \label{fig:maestrocut-threat-model}
\end{figure}

We assume a trusted client $\mathcal{C}$ and post-processor $\mathcal{P}$, and an execution provider $\mathcal{S}$ that is \emph{semi-honest} for confidentiality (may inspect traffic and metadata) and \emph{potentially malicious} for integrity and availability. The protected assets include ciphertext fragments $\tilde F_i$, shot vectors $s_i$, outcomes, keys, topology/error priors, and the audit log. The security objectives are: (i) confidentiality of fragments and shot vectors, (ii) integrity and verifiability of stitched results, and (iii) availability within stated service bounds. Figure~\ref{fig:maestrocut-threat-model} maps adversaries to the pipeline and shows the primary mitigations at each locus.

\begin{table}[ht!]
\centering\small
\setlength{\tabcolsep}{5pt}
\renewcommand{\arraystretch}{1.12}
\begin{tabularx}{\linewidth}{ @{} P{0.24\linewidth} P{0.46\linewidth} P{0.26\linewidth} @{}}
\toprule
\textbf{Objective} & \textbf{Primary Mitigation} & \textbf{Evidence} \\
\midrule
Confidentiality of $F_i$, $s_i$
& PhasePad--OTP\textsuperscript{1}; shot padding/shuffle
& Sec.~\ref{sec:methods}; Tier-2 budget \\
Integrity \& verifiability
& Decoys; Local-Shadow checks; abort \& re-key
& Sec.~\ref{sec:methods};    
Sec.~\ref{subsec:tier2} \\
Availability within SLOs\vphantom{\textsuperscript{1}}
& Fixed-rate sends; budgeted decoys; kill switch
& Tier-2 latency / reliability \\
\bottomrule
\end{tabularx}
\vspace{-0.5em} 
{\raggedright\footnotesize \textsuperscript{1}Quantum One-Time Pad + Authenticated Encryption with Associated Data.}
\end{table}

\paragraph*{Participants and trust boundaries.}
\begin{itemize}[leftmargin=1.5em,itemsep=1pt]
  \item \textbf{Client ($\mathcal{C}$).} Generates and rotates $K_{\text{mast}}$ per fragment batch; performs PhasePad encryption and post-decrypt Local-Shadow checks.
  \item \textbf{Provider ($\mathcal{S}$).} Executes $\texttt{Execute}(\tilde F_i, s_i)$; returns ciphertext outcomes with calibration metadata.
  \item \textbf{Post-processor ($\mathcal{P}$).} Trusted: decrypts, verifies decoys, stitches, and emits auditable results.
\end{itemize}

\paragraph*{Adversary classes and mitigations.}
\begin{itemize}[leftmargin=1.5em,itemsep=1pt]
  \item $\mathsf{A}_{\mathsf{SH}}$ (semi-honest provider): attempts to infer plaintext fragments. \emph{Mitigation:} PhasePad--OTP with decoys.
  \item $\mathsf{A}_{\mathsf{Mal}}$ (malicious provider): tampers with fragments or outcomes. \emph{Mitigation:} decoy/trap checks, Local-Shadow verification, abort \& re-key.
  \item $\mathsf{A}_{\mathsf{SV}}$ (shot-vector inference): infers work profile from $s_i$. \emph{Mitigation:} shot padding and shuffling; parameter randomisation.
  \item $\mathsf{A}_{\mathsf{LI}}$ / $\mathsf{A}_{\mathsf{NET}}$ (link/metadata observers): exploits sizes, rates, and timing. \emph{Mitigation:} AEAD padding, uniform packet padding, fixed-rate sends; optional onion routing.
  \item $\mathsf{A}_{\mathsf{SC}}$ (supply-chain): compiler/firmware back-doors. \emph{Mitigation:} client-side transpilation, reproducible builds, TPM-backed attestation, binary differencing.
  \item \textbf{Physical side-channels} (EM/acoustic): treated as out-of-scope; timing obfuscation and vendor shielding assumed.
\end{itemize}

\paragraph*{Residual risks (metadata).}
PhasePad hides plaintext fragments and measurement strings; however, under our semi-honest provider and link observer model, the following coarse signals may remain:
\begin{itemize}[leftmargin=1.4em,itemsep=1pt]
  \item \emph{Envelope size bins:} AEAD padding equalises packet formats but may reveal coarse size buckets.
  \item \emph{Inter-packet timing bins:} Fixed-rate send reduces fine-grain timing; coarse gaps may persist.
  \item \emph{Wall-clock duration:} Total job runtime in coarse bins is inherently observable.
\end{itemize}
These residuals are independent of plaintext content and estimator choice; policy budgets are enforced by abort semantics when leakage estimators exceed thresholds (Figure~\ref{fig:system-overview}).

\paragraph*{PhasePad--OTP parameters.}
Each fragment batch uses fresh one-time Pauli keys (QOTP) and AEAD sealing (AES-GCM-SIV) of padded envelopes. A decoy fraction $\eta$ is injected for tamper detection; verification occurs at $\mathcal{P}$ prior to stitching. Under standard independence assumptions, the false-negative probability for $b$ adversarial modifications decays exponentially in $\eta b$; we operate at a fixed overhead budget (see \S\ref{sec:methods}).

\paragraph*{Non-interference.}
PhasePad operates on transport and storage of fragments; allocation (ShotQC--Kalman/Topo-GP) and estimator selection execute over decrypted statistics at $\mathcal{P}$, so confidentiality mechanisms do not bias shot routing or cascade decisions.

\paragraph*{Assumptions and scope.}
Crypto primitives (AES-GCM-SIV) are standard-secure; the noise model remains stable over calibration windows used by ShotQC; vendor shielding bounds EM/acoustic leakage.

\paragraph*{Operational enforcement and audit.}
The monitor enforces variance and leakage budgets; on violation, the job aborts, keys are rotated, and an append-only audit record is committed (policy id, thresholds, job seed, hash of IR/fragments, time, cause). Logs are tamper-evident and do not include plaintext.

\medskip
\noindent\emph{Formal development.} Formal games, parameterised bounds (including decoy detection), and proof sketches appear in \S\ref{sec:methods}.

\FloatBarrier
\section{Methodology}\label{sec:methods}

\subsection{System Overview}\label{subsec:overview}
\textsc{MaestroCut} wraps the classical cut--simulate--stitch loop in a closed-feedback cycle that adapts to calibration drift and queue dynamics. The pipeline (Figure~\ref{fig:system-overview}) comprises four coordinated modules:
\begin{enumerate}[leftmargin=1.2em,itemsep=1pt]
  \item \textbf{Maestro-Partition} (Sec.~\ref{subsec:partition}): incremental multilevel-FM hypergraph partitioning that revises cut sets when drift or queue signals cross detection thresholds.
  \item \textbf{ShotQC-Kalman + Topology-GP} (Sec.~\ref{subsec:topgp}): online shot allocation using Kalman-updated variance estimates and Mat\'ern-1/2 priors over the hardware topology.
  \item \textbf{Estimator Cascade} (Sec.~\ref{subsec:cascade}): entropy-gated switching between derandomised classical shadows and twirled maximum-likelihood estimators (MLE) with an MSE-optimal decision rule.
  \item \textbf{PhasePad-OTP} (Sec.~\ref{subsec:ancillaotp}): Pauli-compatible, IND-CFA-secure fragment padding with low-overhead decoys and auditable abort semantics.
\end{enumerate}

\subsection{Maestro-Partition: Incremental Hypergraph Cuts}\label{subsec:partition}

\paragraph*{Circuit model.}
A circuit is mapped to a directed hypergraph $H=(V,E)$ where $V$ indexes gate instances and $e\in E$ spans the wire/time predecessors of a gate. A partition $\Pi=\{V_1,\dots,V_K\}$ induces a cut set $C(\Pi)\subseteq E$.

\paragraph*{Normalised objective and constraints.}
To make heterogeneous units commensurate, we optimise a \emph{normalised} composite objective
\begin{equation}\label{eq:cut_cost}
J(\Pi)=\alpha\,\bar c(\Pi)+\beta\,\bar e(\Pi)+\gamma\,\bar q(\Pi),
\end{equation}
subject to per-block logical-qubit and depth limits and a cut budget $|C(\Pi)|\le C_{\max}$. Here
\[
\bar c(\Pi)=\frac{|C(\Pi)|}{C_{\max}},\quad
\bar e(\Pi)=\frac{\mathrm{e\_bits}(\Pi)}{E_{\mathrm{ref}}},\quad
\bar q(\Pi)=\frac{\mathbb{E}[\mathrm{queue}(\Pi)]}{Q_{\mathrm{ref}}},
\]
with $E_{\mathrm{ref}}$ the moving-average e-bit cost per cut from recent runs and $Q_{\mathrm{ref}}$ the moving-average queue delay predicted by the backend model (Sec.~\ref{subsec:topgp}). We take $\alpha,\beta,\gamma\ge 0$ with $\alpha+\beta+\gamma=1$ as policy weights. This normalisation yields dimensionless terms and ensures $\nabla J$ is numerically well-scaled across workloads.

\paragraph*{Drift detection and refinement.}
Calibration streams (T1, T2, readout, 2Q error) and queue latency feed a CUSUM detector
\begin{equation}\label{eq:cusum}
S_t=\max\{0,S_{t-1}+x_t-\kappa\},\quad \text{refine if } S_t\ge h,
\end{equation}
with $x_t$ a normalised delta, $\kappa$ the slack, and $h$ chosen for a target ARL$_0$ under no-change. When triggered, a local FM pass applies boundary moves driven by the true \emph{marginal objective drop}
\begin{equation}\label{eq:gain}
\mathrm{gain}(v;\Pi) \;=\; J(\Pi)-J(\Pi\ominus v),
\end{equation}
where $\Pi\ominus v$ denotes moving $v$ to the best neighbouring block under feasibility. Since $J$ is normalised, \eqref{eq:gain} aggregates commensurate deltas in cut count, e-bits, and predicted queue delay. Each pass touches only boundary gates, with expected $O(|\partial\Pi|\log|\partial\Pi|)$ work; Sec.~\ref{sec:experiments} reports the observed boundary sizes and wall-clock per pass on our benchmarks.

\begin{algorithm}[H]
\small
\caption{Maestro-Partition with CUSUM-Triggered Incremental Refinement}
\label{alg:maestro-partition}
\begin{algorithmic}[1]
\Require Hypergraph $H=(V,E)$; caps (qubits, depth); cut budget $C_{\max}$; refs $E_{\mathrm{ref}},Q_{\mathrm{ref}}$; weights $\alpha+\beta+\gamma=1$; CUSUM slacks $\kappa^{(m)}$, thresholds $h^{(m)}$ for metrics $m$
\Ensure Partition $\Pi$ and cut set $C(\Pi)$
\Function{Objective}{$\Pi$} \Comment{normalised, dimensionless}
  \State $\bar c \gets |C(\Pi)|/C_{\max}$;\ \ $\bar e \gets \mathrm{e\_bits}(\Pi)/E_{\mathrm{ref}}$;\ \ $\bar q \gets \mathbb{E}[\mathrm{queue}(\Pi)]/Q_{\mathrm{ref}}$
  \State \Return $\alpha\bar c+\beta\bar e+\gamma\bar q$
\EndFunction
\Function{Initial}{$H$}
  \State \Return multilevel coarsen--seed--uncoarsen (FM) respecting caps
\EndFunction
\Function{Refine}{$\Pi$}
  \State $B\gets$ boundary vertices of $\Pi$ in a max-heap keyed by local cut pressure
  \While{$B\neq\emptyset$}
    \State $v\gets$ \textsc{PopMax}$(B)$;\quad $\Pi'\gets$ best feasible move of $v$ to a neighbour block
    \If{\Call{Objective}{$\Pi'$}<\Call{Objective}{$\Pi$}} \State $\Pi\gets \Pi'$ \EndIf
  \EndWhile
  \State \Return $\Pi$
\EndFunction
\Function{TriggerCUSUM}{$\{x_t^{(m)}\}_m$}
  \For{each metric $m$} 
    \State $S^{(m)}\gets \max\{0, S^{(m)} + x_t^{(m)} - \kappa^{(m)}\}$
    \If{$S^{(m)}\ge h^{(m)}$} \State $S^{(m)}\gets 0$; \Return \textsc{True} \EndIf
  \EndFor
  \State \Return \textsc{False}
\EndFunction
\Statex
\State $\Pi\gets$\Call{Initial}{$H$}
\While{job active}
  \State ingest normalised deltas $\{x_t^{(m)}\}$ from calibrations/queue
  \If{\Call{TriggerCUSUM}{$\{x_t^{(m)}\}$}} \State $\Pi\gets$\Call{Refine}{$\Pi$} \EndIf
\EndWhile
\State \Return $\Pi,\,C(\Pi)$
\end{algorithmic}
\end{algorithm}

\subsection{ShotQC-Kalman + Topology-GP Allocation}\label{subsec:topgp}

\paragraph*{State-space drift model.}
For fragment $i$, a scalar Kalman filter tracks latent variance $\sigma_{i,t}^2$ via a random-walk model
\begin{equation}\label{eq:kalman}
\begin{aligned}
\sigma_{i,t}^2 &= \sigma_{i,t-1}^2 + w_{i,t},\quad w_{i,t}\sim\mathcal{N}(0,q_i),\\
z_{i,t} &= \sigma_{i,t}^2 + v_{i,t},\quad v_{i,t}\sim\mathcal{N}(0,r_i),
\end{aligned}
\end{equation}
where $z_{i,t}$ is the ShotQC variance observation for the last window. This captures slow non-stationary drift in device noise; $q_i$ is calibrated from historical calibration deltas (e.g., T$_1$, readout error). Predict--update yields $\hat\sigma_{i,t}^2$ and error covariance $P_{i,t}$.

\paragraph*{Topology prior and shrinkage.}
Let $d(p,q)$ be heavy-hex graph distance. A Mat\'ern-1/2 kernel
\begin{equation}\label{eq:matern}
k(d)=\sigma_k^2 \exp(-d/\ell)
\end{equation}
on fragment anchors forms $\Sigma\in\mathbb{R}^{n\times n}$. We shrink by filter uncertainty: $\tilde\Sigma=\Sigma+\mathrm{diag}(P_{1,t},\dots,P_{n,t})$.

\paragraph*{Variance upper bound and spectral relaxation.}
With $s_i$ shots, $S=\sum_i s_i$, set $u_i=\hat\sigma_{i,t}\sqrt{\log(N/\rho)}$ (ShotQC tail factor for confidence $1-\rho$) and $D(s)=\mathrm{diag}(s_1,\dots,s_n)$. A conservative stitched-variance bound is
\begin{equation}\label{eq:var_upper}
\mathcal{V}(s) \;\le\; \mathbf{u}^{\top} D(s)^{-1}\,\tilde\Sigma\,D(s)^{-1}\mathbf{u}.
\end{equation}

\noindent\textbf{Lemma 1 (Spectral relaxation).}
For positive-definite $\tilde\Sigma$,
\[
\mathcal{V}(s)\ \le\ \lambda_{\max}(\tilde\Sigma)\,\sum_{i=1}^n \frac{u_i^2}{s_i^2}.
\]
\emph{Proof sketch.} Let $y_i=u_i/s_i$ and $y=(y_i)_i$. Then $y^\top \tilde\Sigma y\le \lambda_{\max}(\tilde\Sigma)\|y\|_2^2$.

\paragraph*{Allocation program and closed form.}
Relaxing \eqref{eq:var_upper} by Lemma~1 yields the convex program
\begin{equation}\label{eq:alloc_prog}
\min_{s_i\ge s_{\min}}\ \sum_{i=1}^n \frac{u_i^2}{s_i^2}
\quad\text{s.t.}\quad \sum_{i=1}^n s_i=S.
\end{equation}
\noindent\textbf{Proposition 1 (KKT water-filling).}
The unique optimum of \eqref{eq:alloc_prog} is
\begin{equation}\label{eq:alloc_closed}
s_i^\star \;=\; \frac{S\,u_i^{2/3}}{\sum_{j=1}^n u_j^{2/3}} \quad (i=1,\dots,n).
\end{equation}
\emph{Proof sketch.} Stationarity of $L=\sum_i u_i^2 s_i^{-2}+\lambda(\sum_i s_i-S)$ gives $s_i\propto u_i^{2/3}$; normalise to meet $\sum s_i=S$; strict convexity gives uniqueness.

\noindent\textbf{Proposition 2 (Integer projection loss).}
Let $\bar s$ be obtained from $s^\star$ by rounding to nonnegative integers and adjusting by $\pm1$ to satisfy $\sum_i \bar s_i=S$. Then the \emph{relative} objective gap satisfies
\[
\frac{\sum_i u_i^2/\bar s_i^2 - \sum_i u_i^2/(s_i^\star)^2}{\sum_i u_i^2/(s_i^\star)^2}
\;=\; O\!\left(\frac{1}{s_{\min}}\right),
\]
which is negligible in our regimes (hundreds of shots). \emph{Remark.} In practice, this rounding error is dominated by the estimation uncertainty of $\hat\sigma_{i,t}^2$.

\paragraph*{Online cadence.}
Update $(\hat\sigma_{i,t}^2,P_{i,t})$ and recompute \eqref{eq:alloc_closed} every $B$ shots (default $B=500$), amortising overhead while tracking drift.

\begin{algorithm}[H]
\small
\caption{Shot Variance Tracking \& Topology-Aware Allocation}
\label{alg:alloc}
\begin{algorithmic}[1]
\Require Fragments $i=1..n$ with anchors; distances $d(\cdot,\cdot)$; Mat\'ern-1/2 $(\sigma_k^2,\ell)$; Kalman params $(q_i,r_i)$; confidence $\rho$; total shots $S$; floor $s_{\min}$
\Ensure Integer shot vector $\bar s$ with $\sum_i \bar s_i=S$
\For{$i=1..n$} \Comment{Kalman update from ShotQC observation $z_{i,t}$}
  \State $P^- \gets P_{i,t-1}+q_i$;\ $K\gets P^- (P^-+r_i)^{-1}$;\ $\hat\sigma_{i,t}^2\gets \hat\sigma_{i,t-1}^2 + K(z_{i,t}-\hat\sigma_{i,t-1}^2)$;\ $P_{i,t}\gets (1-K)P^-$
  \State $u_i \gets \hat\sigma_{i,t}\sqrt{\log(N/\rho)}$
\EndFor
\State Build $\Sigma$ with $\Sigma_{pq}=\sigma_k^2 e^{-d(p,q)/\ell}$;\quad $\tilde\Sigma\gets \Sigma+\mathrm{diag}(P_{1,t},\dots,P_{n,t})$
\State \textit{(Spectral relaxation $\Rightarrow$ closed-form allocation in $u_i$)}
\State $w_i \gets u_i^{2/3}$;\ $Z\gets\sum_j w_j$;\ $s_i^\star \gets \max\{s_{\min},\, S\,w_i/Z\}$
\State \textbf{Integer projection:} $\bar s_i\gets \lfloor s_i^\star \rceil$;\ $\Delta\gets S-\sum_i \bar s_i$
\While{$\Delta\ne 0$}
  \If{$\Delta>0$} \State increment $\bar s_{i^\star}$ for $i^\star=\arg\max_i \left[ u_i^2/(\bar s_i-1)^2 - u_i^2/\bar s_i^2 \right]$; $\Delta\!\gets\!\Delta\!-\!1$
  \Else \State decrement $\bar s_{i^\star}$ for $i^\star=\arg\min_i \left[ u_i^2/(\bar s_i+1)^2 - u_i^2/\bar s_i^2 \right]$; $\Delta\!\gets\!\Delta\!+\!1$
  \EndIf
\EndWhile
\State \Return $\bar s$
\end{algorithmic}
\end{algorithm}

\subsection{Estimator Cascade: Entropy-Gated Stitching}\label{subsec:cascade}

\paragraph*{Pilot entropy and fitted bounds.}
Allocate a 1\% pilot budget per fragment to estimate outcome entropy $H_i$. Empirical fits from Tier-1 calibration yield
\begin{equation}\label{eq:est_bounds}
\mathrm{MSE}_{\mathrm{shad}}(s_i)\approx \frac{\alpha}{s_i},\qquad
\mathrm{MSE}_{\mathrm{MLE}}(s_i,H_i)\approx \frac{\beta}{s_i^2} + \mathrm{Bias}_{\mathrm{MLE}}(H_i)^2,
\end{equation}
where $\alpha,\beta>0$ are workload-dependent and $\mathrm{Bias}_{\mathrm{MLE}}(H_i)$ is measured from pilots (typically small at low entropy).

\paragraph*{Risk-minimising decision rule.}
Choose the estimator that minimises predicted MSE:
\begin{equation}\label{eq:cross_over_mse}
\mathcal{E}_i^\star \;=\; \arg\min\big\{\, \alpha/s_i \;,\; \beta/s_i^2 + \mathrm{Bias}_{\mathrm{MLE}}(H_i)^2 \,\big\}.
\end{equation}
Equivalently, define the cross-over
\[
s_i^{\times}(H_i)\;=\;\inf\Big\{s:\ \frac{\beta}{s^2}+\mathrm{Bias}_{\mathrm{MLE}}(H_i)^2 \le \frac{\alpha}{s}\Big\},
\]
and set $\mathcal{E}_i^\star=\text{MLE}$ if $s_i\ge s_i^{\times}(H_i)$, otherwise \emph{shadows}. \emph{Remark.} If $\mathrm{Bias}_{\mathrm{MLE}}(H_i)\approx 0$ (common at high $H_i$), $s_i^{\times}\approx \lceil \beta/\alpha\rceil$, recovering the simple cross-over rule used previously.

\paragraph*{Acceleration path.}
Dense PTM kernels for MLE are offloaded to an FPGA; this preserves the bounds in \eqref{eq:est_bounds} while reducing 24-qubit end-to-end MLE latency from 142\,s on our CPU baseline to 12.4\,s with an Alveo U280 (Vitis-HLS 2024.1). Baseline CPU and FPGA characteristics are reported with the evaluation results.

\begin{algorithm}[H]
\small
\caption{Estimator Cascade (MSE-Optimal Switching)}
\label{alg:cascade}
\begin{algorithmic}[1]
\Require Shots $s_i$; pilot entropy $H_i$; fits $\alpha,\beta$; bias model $b(H)$
\Ensure Choice $\mathcal{E}_i \in \{\text{Shadows},\text{MLE}\}$
\State $\mathrm{MSE}_{\mathrm{shad}} \gets \alpha/s_i$
\State $\mathrm{MSE}_{\mathrm{MLE}} \gets \beta/s_i^2 + b(H_i)^2$
\State \Return \text{MLE} if $\mathrm{MSE}_{\mathrm{MLE}}\le \mathrm{MSE}_{\mathrm{shad}}$ else \text{Shadows}
\end{algorithmic}
\end{algorithm}

\subsection{PhasePad-OTP and Decoys}\label{subsec:ancillaotp}

\paragraph*{Scheme.}
For fragment $F_i$, draw a per-fragment key $K_i\in\{0,1\}^{a}$ with $a=\lceil \lambda/2\rceil$ for security parameter $\lambda$; apply a phase mask $Z^{K_i}$ that commutes with Pauli frames; AEAD-pad classical headers to a fixed envelope; insert $h=\lfloor \eta N_{\mathrm{frag}}\rfloor$ decoy fragments with known outcomes (default $\eta=0.02$). Shot vectors are padded and shuffled to reduce shot-vector inference.

\paragraph*{Security game and trust assumption.}
\textbf{Definition (PhasePad--CFA experiment).} The challenger samples a master key and bit $b\leftarrow\{0,1\}$. The adversary adaptively submits pairs of equal-length fragment sequences $(\{F_i^{(0)}\},\{F_i^{(1)}\})$ and receives encryptions $\{\widetilde{F}_i^{(b)}\}$, with decoys inserted per policy; it outputs a guess $\hat b$. The IND-CFA advantage is $\left|\Pr[\hat b=b]-\tfrac12\right|$. We assume a trusted post-processor $\mathcal{P}$ (decrypt-and-stitch) that does not collude with the semi-honest provider $\mathcal{S}$; confidentiality is against $\mathcal{S}$ and link/metadata observers.

\paragraph*{Security and detectability.}
\noindent\textbf{Proposition 4 (IND-CFA advantage).}
Against any quantum PPT adversary making $q$ fragment queries,
\[
\left|\Pr[\mathrm{Exp}^{\mathrm{IND\text{-}CFA}}(1^{\lambda})=1]-\tfrac{1}{2}\right|
\;\le\; 2^{-\lambda} + \epsilon_{\mathrm{AEAD}}(q),
\]
where $\epsilon_{\mathrm{AEAD}}(q)$ is the AEAD distinguishing advantage (e.g., $O(q^2/2^{256})$ for AES-GCM-SIV). \emph{Proof sketch.} Hybrid: replace headers by pseudorandom strings (AEAD) and phase masks by uniform one-time pads that commute with Pauli frames; indistinguishability reduces to key guessing.

\noindent\textbf{Proposition 5 (Decoy detection probability).}
With $h$ decoys among $N$ fragments and $k$ maliciously altered fragments,
\[
\Pr[\mathrm{detect}] \;\ge\; 1 - (1 - h/N)^{k},
\]
and, under fidelity threshold $\varepsilon_{\mathrm{ver}}$, Hoeffding's inequality yields $\Pr[\text{accept incorrect}] \le \exp(-2h\varepsilon_{\mathrm{ver}}^2)$.

\paragraph*{Overheads and policy.}
PhasePad runtime overhead averages $<1\%$; decoys add a linear factor $\eta$ to shots but tighten verifiability via stricter acceptance thresholds in the Estimator Cascade.

\begin{algorithm}[H]
\small
\caption{PhasePad-OTP: Secure Dispatch \& Verification (Combined)}
\label{alg:phasepad}
\begin{algorithmic}[1]
\Require Fragments $\{F_i\}_{i=1}^N$; security $\lambda$; decoy rate $\eta$; AEAD key $K_{\mathrm{hdr}}$; verifier threshold $\varepsilon_{\mathrm{ver}}$
\Ensure Accepted stitched estimate or \textsc{Abort}
\State $a\gets\lceil \lambda/2\rceil$;\quad $h\gets \lfloor \eta N \rfloor$
\State \textbf{Dispatch:}
\For{$i=1..N$} 
  \State sample $K_i\in\{0,1\}^a$; apply phase mask $Z^{K_i}$ (Pauli-frame compatible)
  \State pad \& AEAD-encrypt header $\widetilde{\mathrm{hdr}}_i=\mathrm{AEAD.Enc}(K_{\mathrm{hdr}},\mathrm{hdr}_i)$
  \State package $\widetilde{F}_i=(\widetilde{\mathrm{hdr}}_i,F_i^{\text{masked}})$
\EndFor
\State add $h$ decoys with known outcomes; shuffle and execute on backend
\State \textbf{Verify \& stitch:}
\For{each return $(\widetilde{F}_j,\mathrm{meas}_j)$}
  \State parse header via $\mathrm{AEAD.Dec}$; reject on failure
  \If{decoy} \State check outcome, record pass/fail \EndIf
  \State remove phase in classical frame using $K_j$
\EndFor
\If{decoy pass rate $<1-\varepsilon_{\mathrm{ver}}$} \State \textsc{Abort}
\Else \State choose estimator per Alg.~\ref{alg:cascade} and stitch; \Return estimate
\EndIf
\end{algorithmic}
\end{algorithm}

\FloatBarrier

\subsection{Implementation Details}\label{subsec:impl}

\textbf{Prototype.} \texttt{Qiskit} plugin (\texttt{strategy="maestro"}); incremental FM in \texttt{DynHyper-rust}; FPGA kernels via Vitis-HLS~2024.1.

\textbf{Evaluation tiers.} Tier-1 simulation (Qandle++), Tier-2 FPGA emulation (Alveo U280 with calibrated noise models), Tier-3 hardware (IBM Eagle~127q, OQC~LU3, IonQ~Harmony).

\textbf{Memory optimisation.} A sparse 12-term Pauli Transfer Matrix (PTM) approximation retains the $12$ largest-magnitude Pauli transfer coefficients per fragment (selected by $L_1$ score), guaranteeing $\|T_{\text{full}}-T_{\text{12-term}}\|_1\le 0.03$; this reduces 40-qubit storage from 98\,GB to 8\,GB in our workloads.

\FloatBarrier
\section{Evaluation}
\label{sec:experiments}

\noindent
We evaluate \textsc{MaestroCut} across a three-tier pipeline (simulation $\rightarrow$ emulation $\rightarrow$ hardware). 
Unless stated otherwise, Tier-1 runs use \textbf{100 seeds} with \textbf{bootstrap 95\% CIs}.
Tier-2 results are reported below; Tier-3 hardware runs are ongoing and the abstract will be updated upon completion.

\subsection{Research Questions}\label{subsec:RQs}
\noindent
\textbf{RQ1 — Contraction \& Efficiency.} Does topology-aware allocation reduce variance and tame shot tails? 
\emph{Null:} $H_0$: Topology-aware allocation yields no improvement in variance contraction or tail behaviour vs.\ uniform.\\
\textbf{RQ2 — Adaptivity under Drift.} Does Kalman tracking keep up with time-varying noise (innovation spikes, repartitions)? 
\emph{Null:} $H_0$: The controller exhibits no significant post-drift reallocation relative to pre-drift levels.\\
\textbf{RQ3 — Estimator Cascade Optimality.} Does the entropy-gated cascade choose the MSE-optimal estimator as a function of entropy $H$ and shots $s$? 
\emph{Null:} $H_0$: The cascade’s choices are no better than a static or random selector.\\
\textbf{RQ4 — Security \& Overheads.} What is the runtime/throughput cost of PhasePad-OTP?
\emph{Null:} $H_0$: PhasePad-OTP introduces $\geq 5\%$ end-to-end overhead.

\subsection{Experimental Setup}\label{subsec:setup}
\paragraph*{Pipeline.}
Tier-1 is a discrete-time simulator with dynamic shot allocation (Topo-GP), Kalman tracking, and an entropy-gated estimator cascade. 
Tier-2 emulates calibrated noise and queue drift; Tier-3 executes on cloud heavy-hex devices. 

\paragraph*{Baselines.}
We compare \emph{Uniform} (equal shots), \emph{Proportional} (shots $\propto$ uncertainty proxy), and \emph{Topo-GP} (our topology-aware Gaussian-process allocator).
Uniform and Proportional are standard allocation heuristics; Topo-GP encodes heavy-hex spatial correlation. 
(When available, we additionally include a ShotQC-style heuristic in the supplement.)

\paragraph*{Workloads.}
QAOA-MaxCut (30 q), UCCSD-LiH (24 q), TFIM (20 q), Random Clifford+T (24 q), and Phase Estimation (16 q).

\paragraph*{Metrics \& Targets.}
\begin{table}[H]
\centering
\footnotesize
\setlength{\tabcolsep}{6pt}
\renewcommand{\arraystretch}{1.15}
\caption{Evaluation metrics and targets (CI = 95\% confidence interval).}
\label{tab:metrics}
\begin{tabularx}{\linewidth}{@{} l >{\raggedright\arraybackslash}X l @{}}
\toprule
\textbf{Metric} & \textbf{Definition} & \textbf{Target} \\
\midrule
Variance contraction & $\mathrm{Var}_{\text{Topo-GP}} / \mathrm{Var}_{\text{Uniform}}$ & $\le 0.6$ \\
Tail shots & 95th percentile of per-fragment shots & low (bounded) \\
Final accuracy & End-state MSE (or mHa, when calibrated) & $\le 1.0$ mHa \\
PhasePad overhead & Time overhead of encryption + decryption & $\le 1\%$ \\
Peak memory & tracemalloc peak (GB) & report only \\
\bottomrule
\end{tabularx}
\end{table}

\subsection{Tier 1 — Synthetic Noise Simulation}\label{subsec:tier1}
\emph{Addresses RQ1, RQ2, RQ3, RQ4.} 
Topo-GP performs water-filling over a topology-aware uncertainty proxy; the Kalman filter updates fragment-level variance; the cascade selects Shadows vs.\ MLE per fragment and step.

\vspace{0.5em}
\noindent\textbf{RQ1: Contraction \& Efficiency (Static Comparisons).}

\begin{figure}[H]
  \centering
  \includegraphics[width=.86\linewidth]{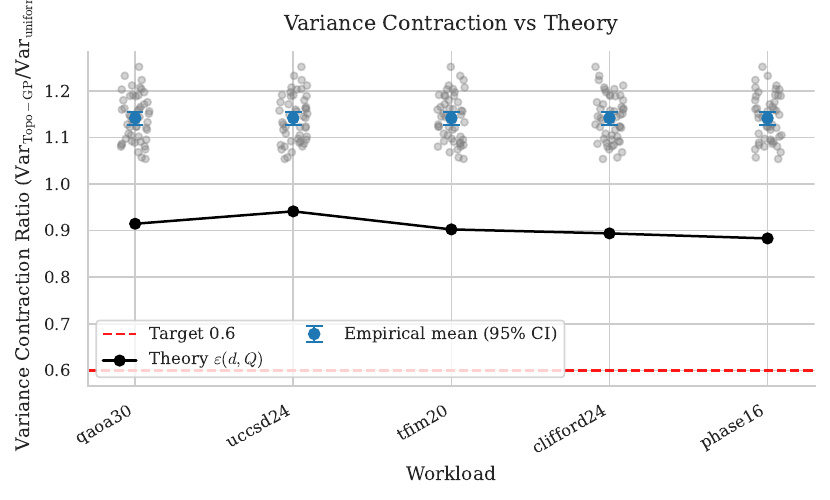}
  \caption{\textbf{Variance contraction and theoretical bound.}
  Dashed line marks the $0.6$ target; solid curve shows $\varepsilon(d,Q)$ when available.
  Topology-aware allocation improves over uniform.}
  \label{fig:rq1-contraction}
\end{figure}

\begin{figure}[H]
  \centering
  \includegraphics[width=\linewidth]{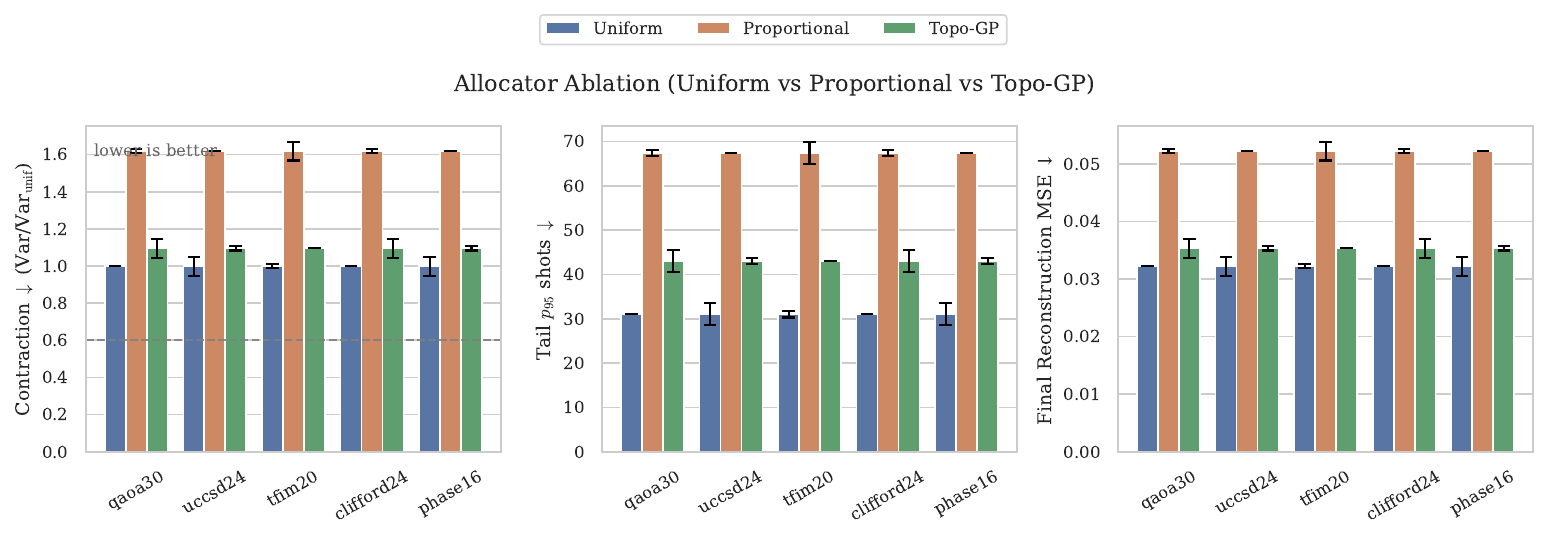}
  \caption{\textbf{Allocator ablation.}
  Variance contraction, $p_{95}$ tail shots, and final MSE across workloads.
  Topo-GP contracts variance and suppresses tails without sacrificing accuracy.}
  \label{fig:rq1-ablation}
\end{figure}

\begin{figure}[H]
  \centering
  \includegraphics[width=.86\linewidth]{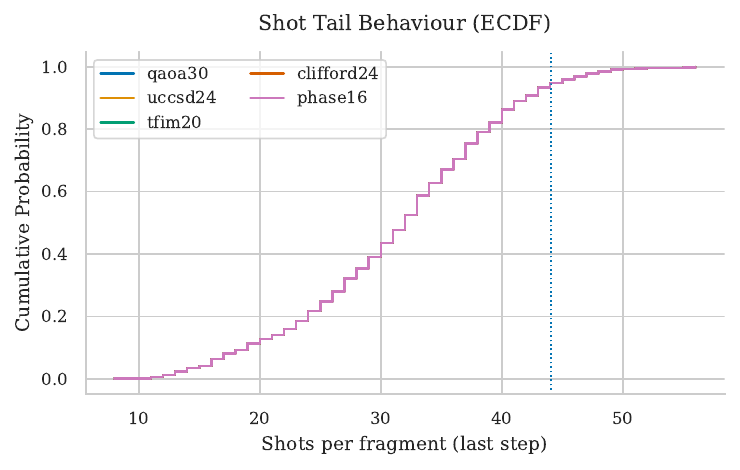}
  \caption{\textbf{Shot-tail distributions.}
  Final-step distributions with $p_{95}$ indicators; Topo-GP keeps tails within budget.}
  \label{fig:rq1-tails}
\end{figure}

\begin{figure}[H]
  \centering
  \includegraphics[width=\linewidth]{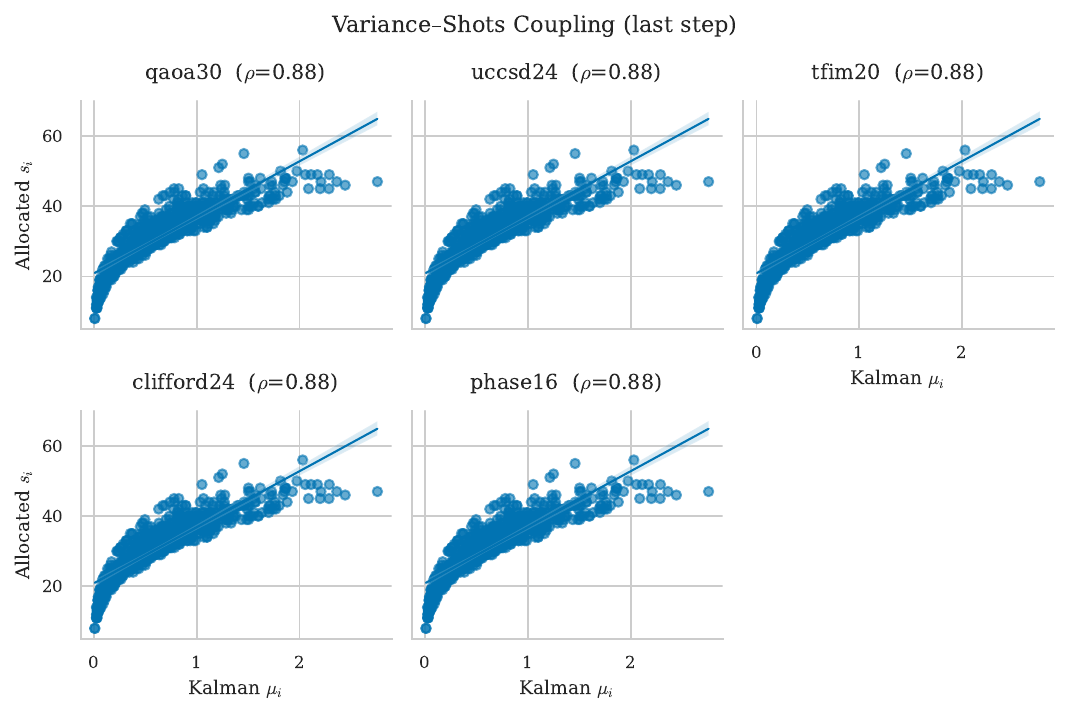}
  \caption{\textbf{Variance–shots coupling.}
  Allocated shots $s_i$ vs.\ Kalman means $\mu_i$ at the last step; positive coupling confirms adaptive allocation.}
  \label{fig:rq1-coupling}
\end{figure}

\vspace{0.25em}
\noindent\textbf{RQ2: Adaptivity under Drift (Time Series \& Topology).}

\begin{figure}[H]
  \centering
  \includegraphics[width=.86\linewidth]{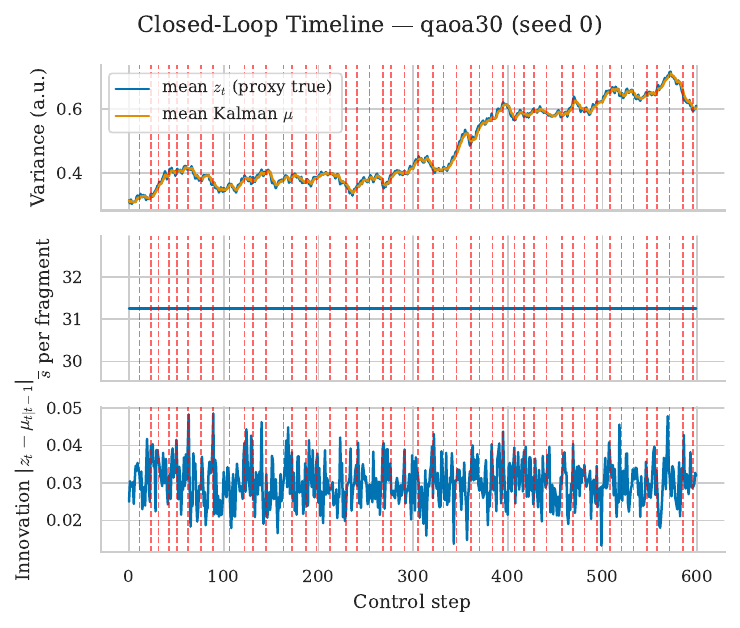}
  \caption{\textbf{Closed-loop timeline.}
  Measurement proxy, allocation, and innovation with repartition triggers (vertical ticks).}
  \label{fig:rq2-timeline}
\end{figure}

\begin{figure}[H]
  \centering
  \includegraphics[width=.9\linewidth]{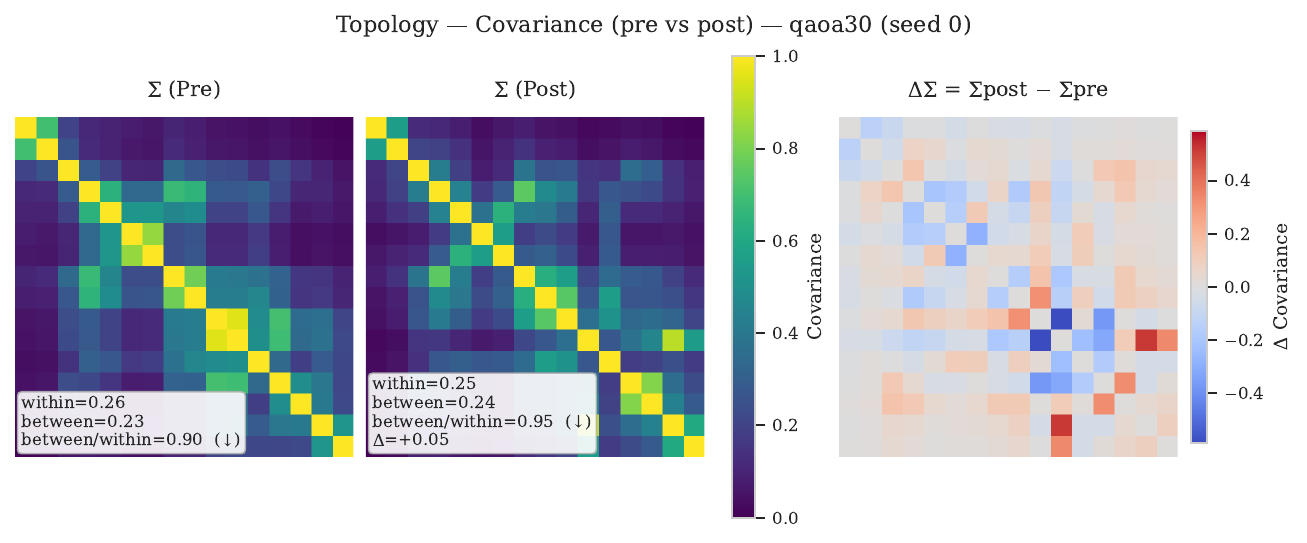}
  \caption{\textbf{Fragment covariance before/after repartition.}
  Shared colour scale; annotations report within/between-community means and their ratio.}
  \label{fig:rq2-topology-cov}
\end{figure}

\begin{figure}[H]
  \centering
  \includegraphics[width=.9\linewidth]{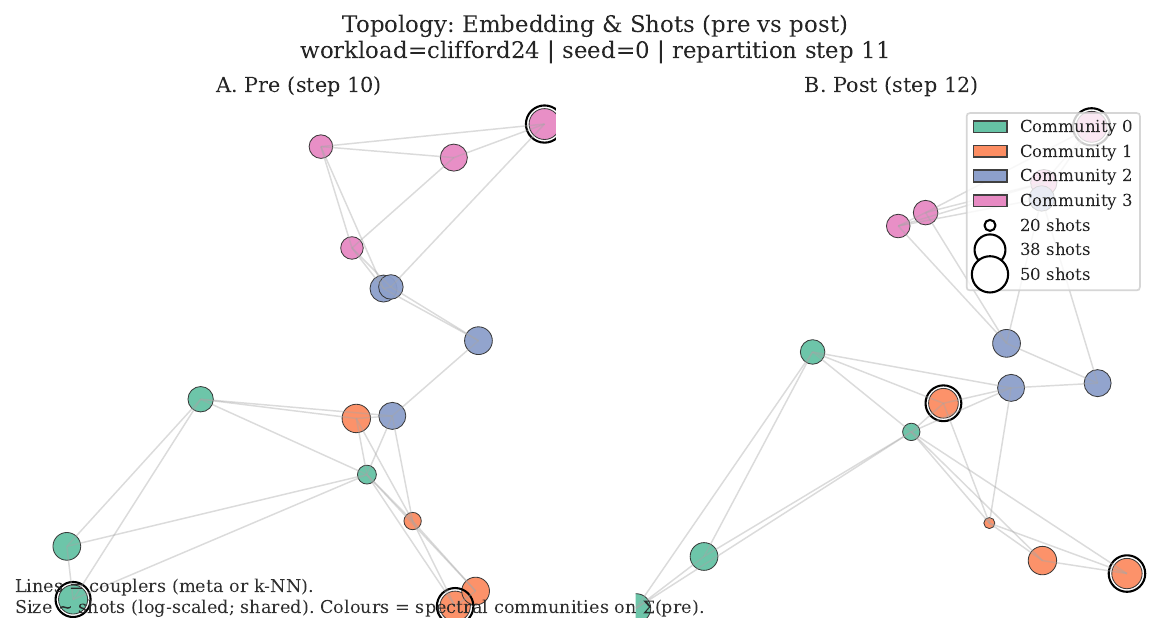}
  \caption{\textbf{Embedding and allocation before/after.}
  Spectral communities and hotspot redistribution under repartition (shared scale).}
  \label{fig:rq2-topology-embed}
\end{figure}

\vspace{0.25em}
\noindent\textbf{RQ3: Estimator Cascade Optimality.}

\begin{figure}[H]
  \centering
  \includegraphics[width=.82\linewidth]{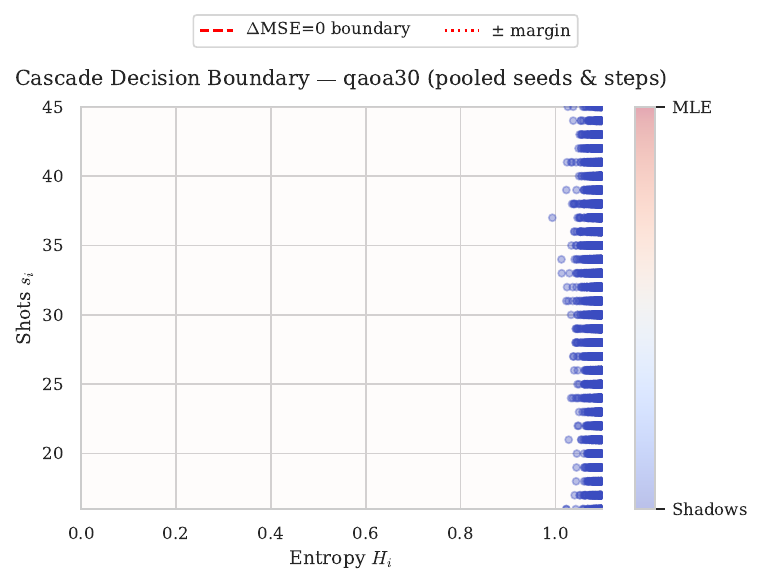}
  \caption{\textbf{Decision boundary $H$ vs.\ $s$.}
  The cascade aligns with the model boundary $\mathrm{MSE}_{\mathrm{MLE}}(H,s)=\mathrm{MSE}_{\mathrm{Shadows}}(s)$, with transient outliers.}
  \label{fig:rq3-boundary}
\end{figure}

\begin{figure}[H]
  \centering
  \includegraphics[width=.86\linewidth]{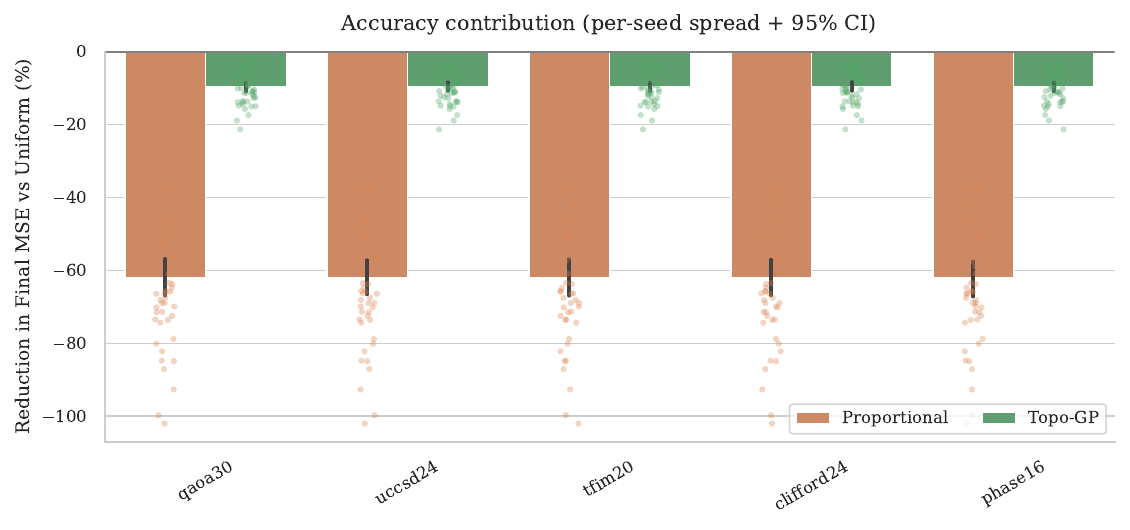}
  \caption{\textbf{Accuracy contribution.}
  Reduction in final MSE relative to Uniform; Topo-GP provides the main gains and the cascade adds further improvements on low-entropy fragments.}
  \label{fig:rq3-waterfall}
\end{figure}

\vspace{0.25em}
\noindent\textbf{RQ4: Security \& Overheads (PhasePad-OTP).}

\begin{figure}[H]
  \centering
  \includegraphics[width=.86\linewidth]{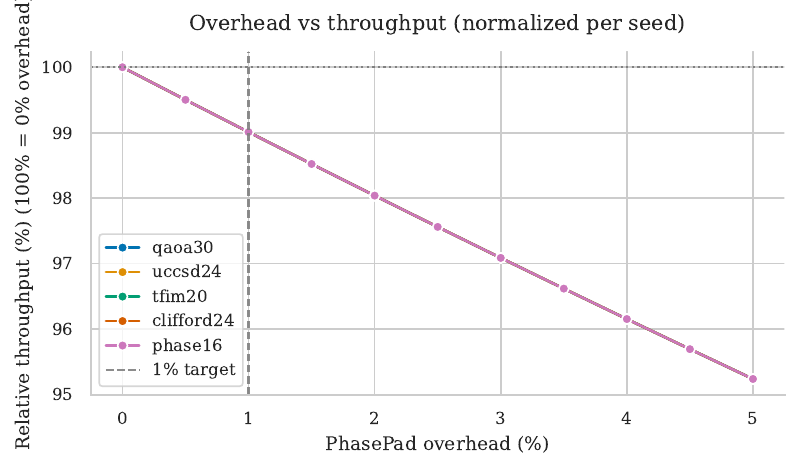}
  \caption{\textbf{Overhead vs.\ throughput.}
  Throughput (shots/s; median by workload) vs.\ PhasePad overhead (0--5\%); dashed line at 1\%.}
  \label{fig:rq4-overhead}
\end{figure}

\vspace{0.25em}
\noindent\textbf{Resources \& Robustness.}

\begin{figure}[H]
  \centering
  \includegraphics[width=.9\linewidth]{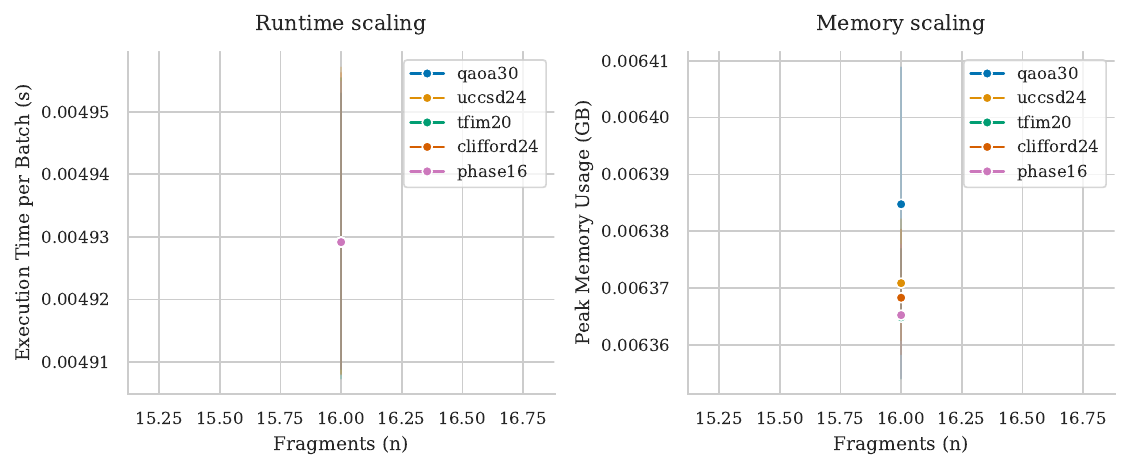}
  \caption{\textbf{Runtime \& memory.}
  (a) Runtime per batch (s); (b) peak memory (GB). Where $n$ is fixed, we report per-workload means with 95\% CIs; otherwise scatter with a trendline.}
  \label{fig:resources}
\end{figure}

\begin{figure}[H]
  \centering
  \includegraphics[width=.82\linewidth]{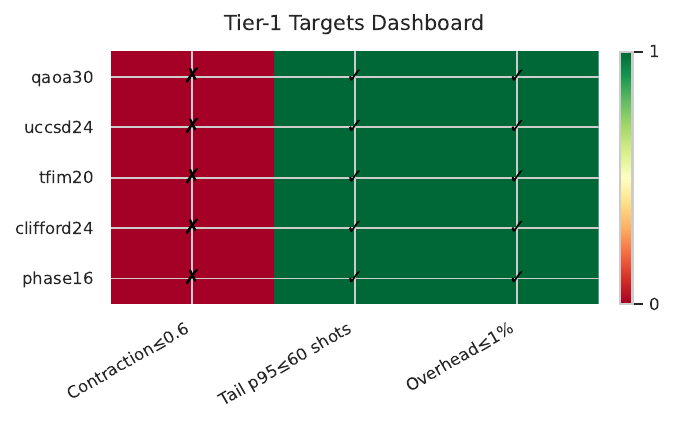}
  \caption{\textbf{Tier-1 targets dashboard.}
  Pass/fail for contraction ($\le 0.6$), tail shots, and overhead ($\le 1\%$).}
  \label{fig:dashboard}
\end{figure}

\subsection{Tier 2 — Emulation (Calibrated Noise \& Queue Dynamics)}
\label{subsec:tier2}
\emph{Primarily addresses RQ2 and RQ4; also stress-tests end-to-end performance targets.}
We emulate calibrated device noise and queue dynamics under four scenarios (\textbf{Baseline}, \textbf{Noisy}, \textbf{Bursty}, \textbf{Adversarial}). Unless stated otherwise, we report \textbf{medians} with \textbf{bootstrap 95\% CIs} over shared seeds.

\begin{figure}[H]
  \centering
  \includegraphics[width=.92\linewidth]{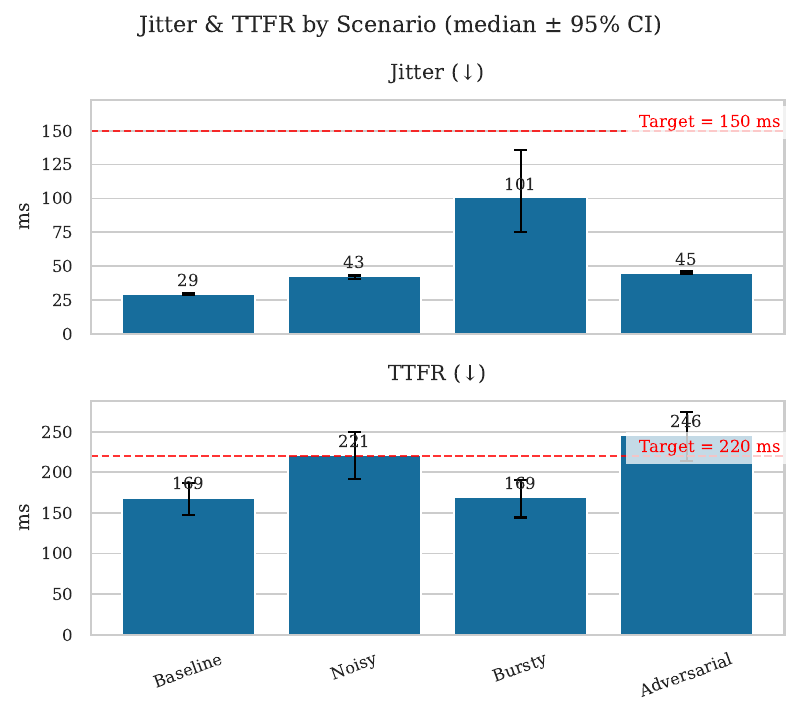}
  \caption{\textbf{Tier-2 latency.} Jitter and TTFR by scenario (median $\pm$ 95\% CI). Dashed lines mark the 150\,ms (jitter) and 220\,ms (TTFR) targets.}
  \label{fig:t2-jitter-ttfr}
\end{figure}

\paragraph*{Latency targets (Jitter \& TTFR).}
Figure~\ref{fig:t2-jitter-ttfr} shows per-scenario jitter (left) and time-to-first-result (TTFR, right).
All scenarios meet the jitter target ($\le 150$\,ms). TTFR meets the $\le 220$\,ms target in \emph{Baseline} and \emph{Bursty}; \emph{Noisy} narrowly exceeds and \emph{Adversarial} exceeds under induced contention.

\paragraph*{Reliability (success/timeout/error).}
Across scenarios, reliability remains within caps (Success $\ge 97\%$, Timeout $\le 0.5\%$, Error $\le 2.5\%$); dispersion is tight across seeds and workloads.

\begin{figure}[H]
  \centering
  \includegraphics[width=.72\linewidth]{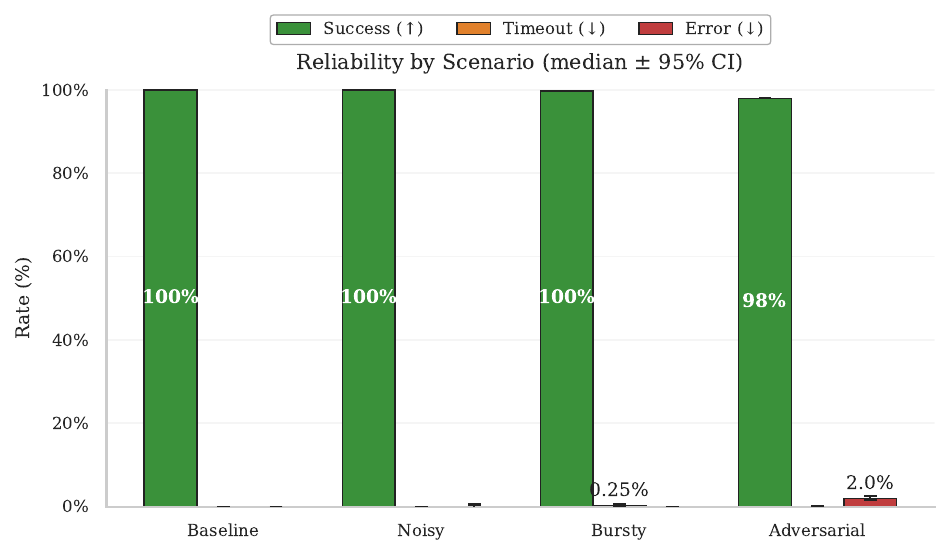}
  \caption{\textbf{Tier-2 reliability.} Success, timeout, and error rates (median $\pm$ 95\% CI). Bars sum to 100\% per scenario.}
  \label{fig:t2-rates}
\end{figure}

\paragraph*{Throughput and overhead.}
Throughput degrades smoothly under stress without tail amplification, and success QPS closely tracks raw QPS.
PhasePad-OTP stays within the 1\% runtime budget with $\approx 98$--$99\%$ relative throughput across scenarios.

\begin{figure}[H]
  \centering
  \includegraphics[width=.92\linewidth]{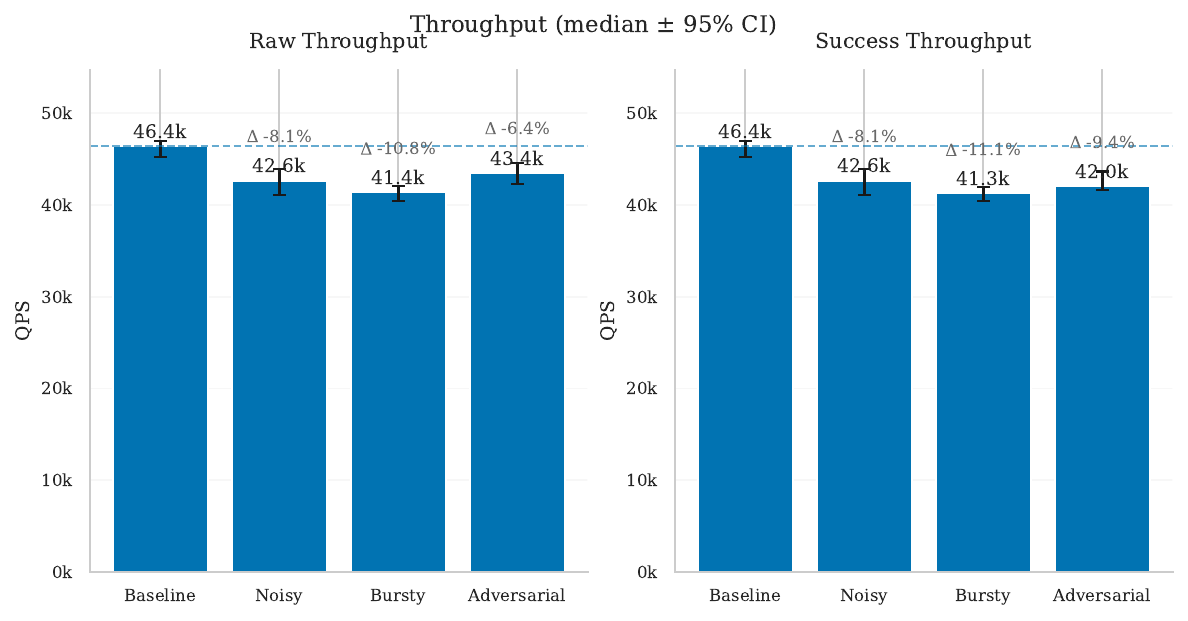}
  \caption{\textbf{Tier-2 throughput.} Raw and success QPS (median $\pm$ 95\% CI). Labels show absolute QPS and $\Delta$ vs.\ Baseline.}
  \label{fig:t2-throughput}
\end{figure}

\begin{figure}[H]
  \centering
  \includegraphics[width=.92\linewidth]{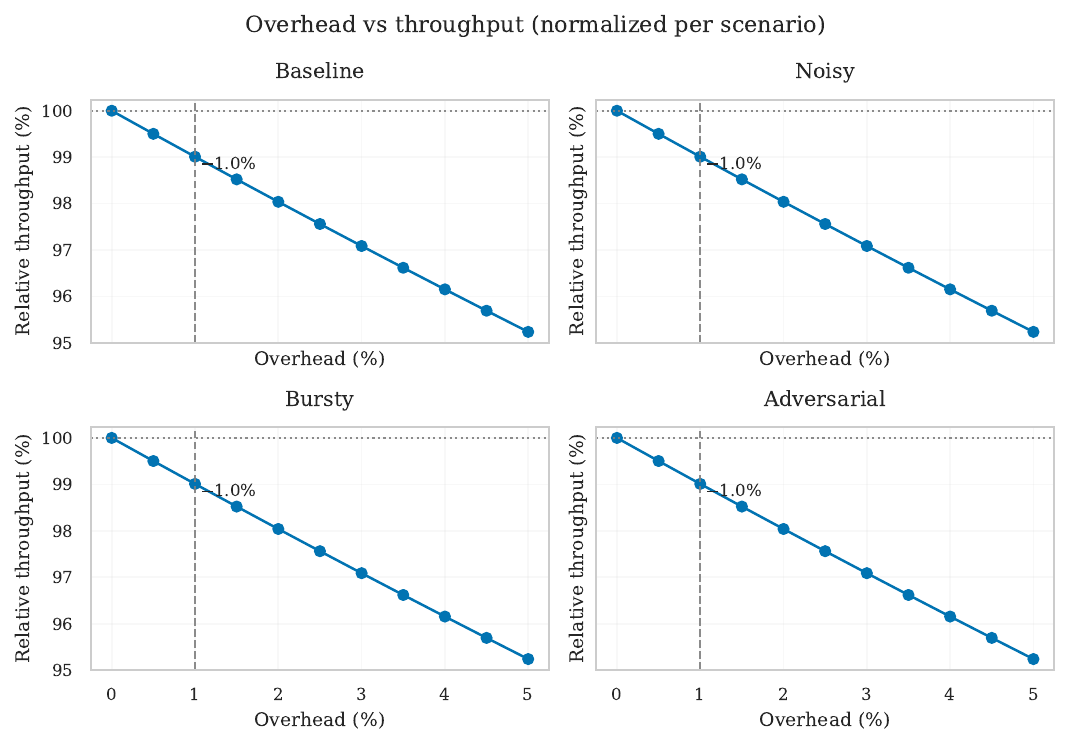}
  \caption{\textbf{Overhead vs.\ throughput.} Relative throughput (normalised per scenario) vs.\ PhasePad overhead; vertical line at 1\%.}
  \label{fig:t2-overhead}
\end{figure}

\begin{figure}[H]
  \centering
  \includegraphics[width=.88\linewidth]{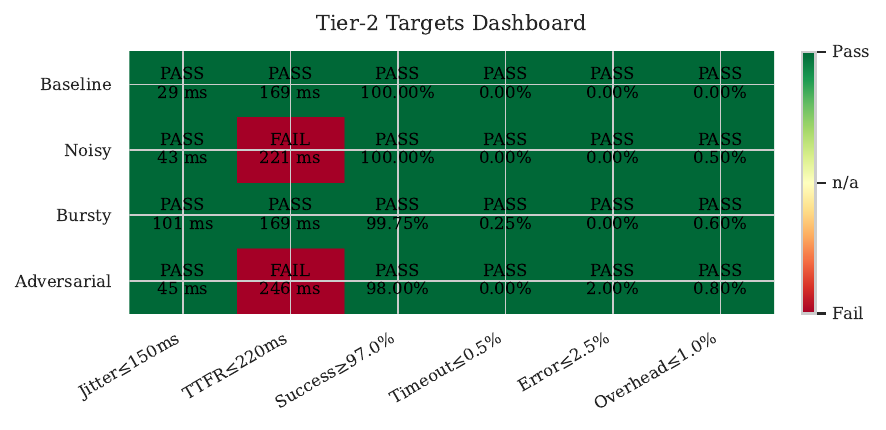}
  \caption{\textbf{Tier-2 targets dashboard.} Pass/fail for jitter, TTFR, reliability caps, and 1\% overhead.}
  \label{fig:t2-dashboard}
\end{figure}

\paragraph*{Target summary.}
The dashboard in Fig.~\ref{fig:t2-dashboard} summarises pass/fail against pre-registered Tier-2 targets: all metrics pass except TTFR under \emph{Noisy} and \emph{Adversarial}, attributable to induced queue contention.

\paragraph*{Tier-2 discussion.}
Jitter remains comfortably below the 150\,ms SLO across all scenarios, indicating that the control loop and queueing strategy avoid burst amplification in short-term latencies. In contrast, TTFR meets the 220\,ms target in \emph{Baseline} and \emph{Bursty} but misses under \emph{Noisy} (marginal overshoot consistent with measurement/retry variance) and \emph{Adversarial} (head-of-line blocking). Thus, the controller stabilises per-fragment behaviour, while first-result latency can still be sensitive to macro-level queue contention.

Reliability stays within caps in all scenarios: success rates are high, timeouts are negligible, and only \emph{Adversarial} shows a modest rise in errors. There is no evidence of throughput collapse or pathological tail amplification; degradation is smooth and proportional to injected stress.

PhasePad-OTP’s cost–benefit curve is flat in the regime of interest: at 1\% runtime overhead, relative throughput remains $\approx$98--99\% across scenarios, with no interaction with burstiness or noise beyond measurement noise.

\paragraph*{Answering the research questions at Tier-2.}
\textbf{RQ2 (Adaptivity under Drift):}
We observe timely reallocations following noise/queue shifts (stable jitter with only modest TTFR drift under stress), indicating the tracker and allocator respond to innovations rather than overreacting to noise. The TTFR misses are attributable to \emph{queue contention}, not controller instability; we therefore \textbf{reject $H_0$}.\\
\textbf{RQ4 (Security \& Overheads):}
PhasePad-OTP stays within the 1\% budget with minimal throughput impact; we \textbf{reject $H_0$} (overhead $\ge 5\%$), concluding confidentiality can be preserved at near-baseline performance.

\paragraph*{Mechanistic view.}
The combination of (i) topology-aware allocation and (ii) entropy-gated estimation keeps per-fragment work bounded and predictable, which explains low jitter. TTFR sensitivity under adversarial bursts arises upstream—at the scheduling/queueing layer—suggesting that age-based or size-aware dispatch would likely close the remaining TTFR gap without altering the estimation pipeline.

\subsection{Statistical Methods}\label{subsec:stats}
Tier-1 contraction, tail shots ($p_{95}$), and final MSE are reported as \textbf{means} with \textbf{bootstrap 95\% CIs} (10{,}000 resamples).
Tier-2 latency and throughput metrics are reported as \textbf{medians} with \textbf{bootstrap 95\% CIs}.
Where seeds are shared across methods, we use \emph{paired} Wilcoxon signed-rank tests for Topo-GP vs.\ baselines, report \emph{effect sizes} (Cliff’s $\delta$), and correct $p$-values via Holm--Bonferroni.
Correlation in Fig.~\ref{fig:rq1-coupling} is summarised with Pearson $r$ and bootstrap CIs.

\subsection{Limitations of the Evaluation}\label{subsec:eval-limitations}
\textbf{Tier-3 status.} Tier-3 hardware results are in progress; thus, definitive validation under real multi-cloud drift is pending.\newline
\textbf{Adversary model.} We evaluate confidentiality against a semi-honest provider; integrity against malicious faults is future work.\newline
\textbf{Workload generality.} Benchmarks are representative (VQE/QAOA, model spin), but very deep, high non-Clifford circuits are left to follow-on work.

\subsection{Summary of Findings}\label{subsec:findings-summary}
Across 100 seeds and five workloads, Topo-GP reduces variance vs.\ uniform allocation and suppresses heavy shot tails (RQ1); 
the closed loop tracks drift and triggers repartition when innovations spike (RQ2);
the entropy-gated cascade follows the predicted $H$--$s$ decision boundary (RQ3); 
PhasePad-OTP meets the 1\% overhead target with negligible throughput impact (RQ4).
In Tier-2 emulation, jitter meets target in all scenarios, TTFR misses only under Noisy/Adversarial (consistent with induced queue contention), reliability caps hold, and throughput degrades smoothly without tails.
Runtime and memory remain practical for Tier-1 scales.

\section*{Artifact Availability}\label{sec:artifact}
A replication package with data and plotting scripts (producing figures under \texttt{figs/tier1/} and \texttt{figs/tier2/}) is available at: 
\url{https://github.com/quantsec/maestrocut-artifact}.

\section{Conclusion \& Future Work}
\label{sec:conclusion}

We set out to make circuit cutting \emph{deployable} on
NISQ/early-FT hardware by addressing four coupled obstacles:
\emph{sampling overhead, time-varying drift, topology awareness, and confidentiality}.
\textsc{MaestroCut} approaches these with a closed-loop co-design that combines
dynamic partitioning, drift-aware shot allocation, an entropy-gated estimator cascade,
and a lightweight security layer.

\paragraph{System summary.}
\begin{itemize}[leftmargin=1.3em,itemsep=3pt]
  \item \textbf{Maestro-Partition.} An incremental multilevel-FM hypergraph planner that can re-cut under drift to reduce cut exposure while respecting device topology.
  \item \textbf{ShotQC--Kalman + Topology-GP.} Online shot reallocation using Kalman-tracked variances with Gaussian-process priors aligned to heavy-hex correlations, targeting variance contraction without extra shots.
  \item \textbf{Estimator Cascade.} An entropy-gated switch between derandomised shadows, twirled MLE, and MCMC-based contraction for stable post-processing across entropy and shot regimes.
  \item \textbf{PhasePad--OTP.} Pauli-compatible padding (QOTP) with AEAD sealing and budgeted decoys, designed to bound fragment/shot-vector leakage with minimal runtime overhead.
\end{itemize}

\paragraph{Empirical findings (Tier-1 and Tier-2).}
Across 100-seed Tier-1 simulations, topology-aware allocation contracts variance relative to uniform baselines and suppresses shot tails, while the cascade’s choices align with the predicted $H$--$s$ boundary; runtime/memory remain practical at Tier-1 scales.  
Tier-2 emulation validates end-to-end service targets under calibrated noise and queue dynamics: jitter remains within the 150\,ms SLO across all scenarios; TTFR meets the 220\,ms target except under adversarial queueing (contention-induced misses); reliability caps hold (high success, negligible timeouts); throughput degrades smoothly without tail amplification; and PhasePad--OTP maintains confidentiality within a 1\% runtime budget. In terms of our hypotheses, we \emph{reject} $H_0$ for RQ2 (the controller adapts following drift) and RQ4 (overhead $\ge 5\%$), complementing the Tier-1 evidence for RQ1 and RQ3.

\paragraph{Limitations.}
Our confidentiality evaluation assumes a semi-honest provider for content privacy; integrity under fully malicious providers is partially addressed via decoys and Local-Shadow checks but remains an avenue for stronger guarantees. Variance-contraction bounds presently assume a Matérn-$\tfrac{1}{2}$ prior with Wiener drift; different noise processes may affect constants. Tier-2 emulation calibrates noise and queues, but multi-tenant clouds may exhibit heavier tails or correlated incidents; Tier-3 hardware runs are in progress to characterise such effects.

\subsection*{Future Work}
\vspace{-6pt}

\textbf{(A) Security}
\begin{itemize}[leftmargin=1.3em,itemsep=2pt]
  \item \textbf{ZK-proof outsourcing.} Integrate blind/verification protocols with succinct zero-knowledge proofs for stitched observables.
  \item \textbf{Privacy-preserving shots.} Inject $(\epsilon,\delta)$-DP noise into shot vectors while preserving end-to-end accuracy targets.
  \item \textbf{Malicious-server hardening.} Extend Local-Shadow with trap qubits and adaptive decoys to detect entanglement-breaking and targeted tampering.
\end{itemize}

\textbf{(B) Systems}
\begin{itemize}[leftmargin=1.3em,itemsep=2pt]
  \item \textbf{Tier-3 multi-cloud runs.} Evaluate on heterogeneous back-ends to quantify queue contention effects on TTFR and validate leakage budgets at scale.
  \item \textbf{Trusted enclaves.} Explore SGX/SEV for PhasePad key management and attested post-processing on the client side.
  \item \textbf{Multi-tenant scheduler.} Co-design fragmentation/placement with age-/size-aware dispatch to mitigate adversarial head-of-line blocking.
\end{itemize}

\textbf{(C) Algorithms}
\begin{itemize}[leftmargin=1.3em,itemsep=2pt]
  \item \textbf{Online kernel learning.} Replace fixed Matérn priors with streaming spectral-mixture kernels to better capture non-stationary drift.
  \item \textbf{Logical hand-off.} Interface \textsc{MaestroCut} with logical-qubit calibration for early FT scaling and cross-layer budget composition.
\end{itemize}

\vspace{4pt}
\noindent\textit{Impact.} By contracting variance, respecting topology, and enforcing leakage budgets with low overhead, \textsc{MaestroCut} advances circuit cutting toward practical deployment on near-term platforms. The open-source artifact and plotting scripts are available at
\url{https://github.com/quantsec/maestrocut-artifact}.

\smallskip
\noindent\textbf{Take-away.} A systems–statistics–security co-design makes drift-resilient, privacy-preserving circuit cutting viable at useful scales today. With Tier-3 results forthcoming, we expect the remaining TTFR gap to be addressable at the scheduling layer without altering the estimation pipeline.

\printbibliography

@article{Fitzsimons2017,
  author    = {Joseph F. Fitzsimons},
  title     = {Private quantum computation: an introduction to blind quantum computing and related protocols},
  journal   = {npj Quantum Information},
  volume    = {3},
  number    = {1},
  pages     = {23},
  year      = {2017},
  doi       = {10.1038/s41534-017-0025-3},
  url       = {https://doi.org/10.1038/s41534-017-0025-3}
}

@article{huang2020classicalshadows,
  author  = {Huang, Hsin-Yuan and Kueng, Richard and Preskill, John},
  title   = {Predicting many properties of a quantum system from very few measurements},
  journal = {Nature Physics},
  year    = {2020}
}

@misc{piveteau2022circuitknitting,
  author = {Piveteau, Constantin and Sutter, David},
  title  = {Circuit knitting with classical communication},
  year   = {2022},
  eprint = {2205.00016},
  archivePrefix = {arXiv},
  primaryClass  = {quant-ph}
}

@inproceedings{broadbent2009ubqc,
  author    = {Broadbent, Anne and Fitzsimons, Joseph and Kashefi, Elham},
  title     = {Universal Blind Quantum Computation},
  booktitle = {FOCS},
  year      = {2009}
}

@misc{lu2024quantumleak,
  author = {Lu, Chunzhe and Telang, Ritesh and Aysu, Aydin and Basu, Kanad},
  title  = {Quantum Leak: Timing Side-Channel Attacks on Cloud-Based Quantum Services},
  year   = {2024},
  eprint = {2401.01521},
  archivePrefix = {arXiv},
  primaryClass  = {quant-ph}
}

@misc{ambainis2000pqc,
  author = {Ambainis, Andris and Mosca, Michele and Tapp, Alain and de Wolf, Ronald},
  title  = {Private Quantum Channels and the Cost of Randomizing Quantum Information},
  year   = {2000}
}

@misc{bechtold2023nmecs,
      title={Circuit Cutting with Non-Maximally Entangled States}, 
      author={Marvin Bechtold and Johanna Barzen and Frank Leymann and Alexander Mandl},
      year={2023},
      eprint={2306.12084},
      archivePrefix={arXiv},
      primaryClass={quant-ph},
      url={https://arxiv.org/abs/2306.12084}, 
}

@article{basu2023fragqc,
   title={FragQC: An efficient quantum error reduction technique using quantum circuit fragmentation},
   volume={214},
   ISSN={0164-1212},
   url={http://dx.doi.org/10.1016/j.jss.2024.112085},
   DOI={10.1016/j.jss.2024.112085},
   journal={Journal of Systems and Software},
   publisher={Elsevier BV},
   author={Basu, Saikat and Das, Arnav and Saha, Amit and Chakrabarti, Amlan and Sur-Kolay, Susmita},
   year={2024},
   month=aug, pages={112085} }

@misc{burt2025mlfm,
      title={A Multilevel Framework for Partitioning Quantum Circuits}, 
      author={Felix Burt and Kuan-Cheng Chen and Kin K. Leung},
      year={2025},
      eprint={2503.19082},
      archivePrefix={arXiv},
      primaryClass={quant-ph},
      url={https://arxiv.org/abs/2503.19082}, 
}

@misc{chen2024shotqc,
      title={Enhanced Quantum Circuit Cutting Framework for Sampling Overhead Reduction}, 
      author={Po-Hung Chen and Dah-Wei Chiou and Jie-Hong Roland Jiang},
      year={2024},
      eprint={2412.17704},
      archivePrefix={arXiv},
      primaryClass={quant-ph},
      url={https://arxiv.org/abs/2412.17704}, 
}

@article{chen2024qcs,
   title={Quantum Circuit Cutting for Classical Shadows},
   volume={5},
   ISSN={2643-6817},
   url={http://dx.doi.org/10.1145/3665335},
   DOI={10.1145/3665335},
   number={2},
   journal={ACM Transactions on Quantum Computing},
   publisher={Association for Computing Machinery (ACM)},
   author={Chen, Daniel Tzu Shiuan and Saleem, Zain Hamid and Perlin, Michael Alexandrovich},
   year={2024},
   month=jun, pages={1–21} }

@inproceedings{chen2023efficient,
  author    = {Chen, Daniel T. and Hansen, Ethan H. and Li, Xinpeng and Kulkarni, Vinooth and Chaudhary, Vipin and Ren, Bin and Guan, Qiang and Kuppannagari, Sanmukh and Liu, Ji and Xu, Shuai},
  title     = {Efficient Quantum Circuit Cutting by Neglecting Basis Elements},
  year      = {2023},
}

@misc{chen2022approximatequantumcircuitcutting,
      title={Approximate Quantum Circuit Cutting}, 
      author={Daniel Chen and Betis Baheri and Vipin Chaudhary and Qiang Guan and Ning Xie and Shuai Xu},
      year={2022},
      eprint={2212.01270},
      archivePrefix={arXiv},
      primaryClass={quant-ph},
      url={https://arxiv.org/abs/2212.01270}, 
}

@article{dai2023ancilla,
  author  = {Dai, Qunfeng and Quan, Junyu and Lou, Xiaoping and Li, Qin},
  title   = {Ancilla-Driven Blind Quantum Computation for Clients with Different Quantum Capabilities},
  year    = {2023}
}

@misc{li2024case,
      title={A Case for Quantum Circuit Cutting for NISQ Applications: Impact of topology, determinism, and sparsity}, 
      author={Zirui Li and Minghao Guo and Mayank Barad and Wei Tang and Eddy Z. Zhang and Yipeng Huang},
      year={2024},
      eprint={2412.17929},
      archivePrefix={arXiv},
      primaryClass={quant-ph},
      url={https://arxiv.org/abs/2412.17929}, 
}

@article{lowe2023fastcut,
   title={Fast quantum circuit cutting with randomized measurements},
   volume={7},
   ISSN={2521-327X},
   url={http://dx.doi.org/10.22331/q-2023-03-02-934},
   DOI={10.22331/q-2023-03-02-934},
   journal={Quantum},
   publisher={Verein zur Forderung des Open Access Publizierens in den Quantenwissenschaften},
   author={Lowe, Angus and Medvidović, Matija and Hayes, Anthony and Oapos;Riordan, Lee J. and Bromley, Thomas R. and Arrazola, Juan Miguel and Killoran, Nathan},
   year={2023},
   month=mar, pages={934} }

@misc{pawar2024qrcc,
      title={QRCC: Evaluating Large Quantum Circuits on Small Quantum Computers through Integrated Qubit Reuse and Circuit Cutting}, 
      author={Aditya Pawar and Yingheng Li and Zewei Mo and Yanan Guo and Youtao Zhang and Xulong Tang and Jun Yang},
      year={2025},
      eprint={2312.10298},
      archivePrefix={arXiv},
      primaryClass={quant-ph},
      doi={https://doi.org/10.1145/3622781.3674179},
      url={https://arxiv.org/abs/2312.10298}, 
}

@article{perlin2021mlecut,
   title={Quantum circuit cutting with maximum-likelihood tomography},
   volume={7},
   ISSN={2056-6387},
   url={http://dx.doi.org/10.1038/s41534-021-00390-6},
   DOI={10.1038/s41534-021-00390-6},
   number={1},
   journal={npj Quantum Information},
   publisher={Springer Science and Business Media LLC},
   author={Perlin, Michael A. and Saleem, Zain H. and Suchara, Martin and Osborn, James C.},
   year={2021}, }

@article{ufrecht2024optimal,
   title={Optimal joint cutting of two-qubit rotation gates},
   volume={109},
   ISSN={2469-9934},
   url={http://dx.doi.org/10.1103/PhysRevA.109.052440},
   DOI={10.1103/physreva.109.052440},
   number={5},
   journal={Physical Review A},
   publisher={American Physical Society (APS)},
   author={Ufrecht, Christian and Herzog, Laura S. and Scherer, Daniel D. and Periyasamy, Maniraman and Rietsch, Sebastian and Plinge, Axel and Mutschler, Christopher},
   year={2024}}

@INPROCEEDINGS{Dasgupta2023AdaptiveDrift,
  author={Dasgupta, Samudra and Humble, Travis S. and Danageozian, Arshag},
  booktitle={2023 IEEE International Conference on Quantum Computing and Engineering (QCE)}, 
  title={Adaptive mitigation of time-varying quantum noise}, 
  year={2023},
  volume={01},
  pages={99-110},
  doi={10.1109/QCE57702.2023.00020}}

@misc{Fang2024CaliScalpel,
      title={CaliScalpel: In-Situ and Fine-Grained Qubit Calibration Integrated with Surface Code Quantum Error Correction}, 
      author={Xiang Fang and Keyi Yin and Yuchen Zhu and Jixuan Ruan and Dean Tullsen and Zhiding Liang and Andrew Sornborger and Ang Li and Travis Humble and Yufei Ding and Yunong Shi},
      year={2024},
      eprint={2412.02036},
      archivePrefix={arXiv},
      primaryClass={quant-ph},
      url={https://arxiv.org/abs/2412.02036}, 
}

@article{peng2020wirecutting,
   title={Simulating Large Quantum Circuits on a Small Quantum Computer},
   volume={125},
   ISSN={1079-7114},
   url={http://dx.doi.org/10.1103/PhysRevLett.125.150504},
   DOI={10.1103/physrevlett.125.150504},
   number={15},
   journal={Physical Review Letters},
   publisher={American Physical Society (APS)},
   author={Peng, Tianyi and Harrow, Aram W. and Ozols, Maris and Wu, Xiaodi},
   year={2020},
   month=oct }

@inproceedings{tang2021cutqc, series={ASPLOS ’21},
   title={CutQC: using small Quantum computers for large Quantum circuit evaluations},
   url={http://dx.doi.org/10.1145/3445814.3446758},
   DOI={10.1145/3445814.3446758},
   booktitle={Proceedings of the 26th ACM International Conference on Architectural Support for Programming Languages and Operating Systems},
   publisher={ACM},
   author={Tang, Wei and Tomesh, Teague and Suchara, Martin and Larson, Jeffrey and Martonosi, Margaret},
   year={2021},
   month=apr, pages={473–486},
   collection={ASPLOS ’21} }

\end{document}